\documentclass [reprint aps,prb,twocolumn,groupedaddress,nofootinbib,floatfix,amsmath,mssymb,prper,floatfix,twocolumn, superscriptaddress]{revtex4-2}
\usepackage{array, makecell}
\usepackage{graphicx} 
\graphicspath{{./Graphics/}}
\usepackage{graphics}
\usepackage{epsfig}
\usepackage{dcolumn}
\usepackage{comment}
\usepackage{amsfonts,color}
\usepackage{appendix}
\usepackage[utf8]{inputenc}
\usepackage{bm}
\usepackage{scalerel,amssymb}

\begin{document}
\preprint{APS/123-QED}
\title{Challenges and outcomes in remote undergraduate research programs during the COVID-19 pandemic}
%Who says research can't be done remotely?}%: Reality of the Remote Undergraduate Research Experience During the COVID-19 Pandemic}
\author{Dina Zohrabi Alaee} \author{Benjamin M. Zwickl}
\affiliation{%
School of Physics and Astronomy, Rochester Institute of Technology, 84 Lomb Memorial Drive, Rochester, NY,14623}%
\date{\today}
\begin{abstract}
In the Summer of 2020, as COVID-19 limited in-person research opportunities and created additional barriers for many students, institutions either canceled or remotely hosted their Research Experience for Undergraduates (REU) programs. The present qualitative phenomenographic study was designed to explore some of the possible limitations, challenges and outcomes of this remote experience. Overall, 94 interviews were conducted with paired participants; mentees ($N$=10) and mentors ($N$=8) from six different REU programs. By drawing on Cultural-Historical Activity Theory (CHAT) as a framework, our study uncovers some of the challenges for mentees around the accomplishment of their research objectives and academic goals. These challenges included motivation, limited access to technologies at home, limited communication among REU students, barriers in mentor-mentee relationships, and differing expectations about doing research. Despite the challenges, all mentees reported that this experience was highly beneficial. Comparisons between outcomes of these remote REUs and published outcomes of in-person undergraduate research programs reveal many similar benefits such as integrating students into STEM culture. Our study suggests that remote research programs could be considered as a means to expand access to research experiences for undergraduate students even after COVID-19 restrictions have been lifted.
\end{abstract}
\maketitle
\section{Introduction}
Undergraduate research experiences (UREs) affect students' academic pathways and career preparation towards STEM by providing authentic research-based learning situations \cite{council_transforming_1999, council_improving_2002, wenzel_enhancing_2004, Kenny_2001, kuh_high-impact_2008}. A large body of literature has reported both academic and psychosocial benefits as outcomes of in-person UREs. Academically, UREs have been reported to help students achieve a higher level of content knowledge \cite{kardash_evaluation_2000}, while also improving their eventual career outcomes \cite{lopatto_undergraduate_2007, hathaway_relationship_2002}. Meanwhile, psychosocial benefits refer to the positive growth in a student's perceptions, emotions, attitudes, and social dimensions around their academic experiences. Psychosocially, in-person UREs have been shown to help students increase self-confidence \cite{seymour_establishing_2004, lopatto_survey_2004, hunter_becoming_2007, laursen_undergraduate_2010}, develop communication skills \cite{kardash_evaluation_2000, hunter_becoming_2007}, improve scientific identity \cite{lopatto_undergraduate_2007,estrada_toward_2011, graham_increasing_2013}, and grow a sense of belonging to the community \cite{lopatto_undergraduate_2007, hausmann_sense_2007, dolan_toward_2009, eagan_making_2013}.

One structure of a URE, which is common in the United States, is the Research Experiences for Undergraduates (REU) program, which is a ten-week summer research experience funded by the U.S. National Science Foundation. Due to the COVID-19 pandemic, some REU programs transitioned to a remote format in the summer of 2020.

\begin{figure}[h]
\centering
\includegraphics [trim=267 155 190 130,clip,width=85mm]{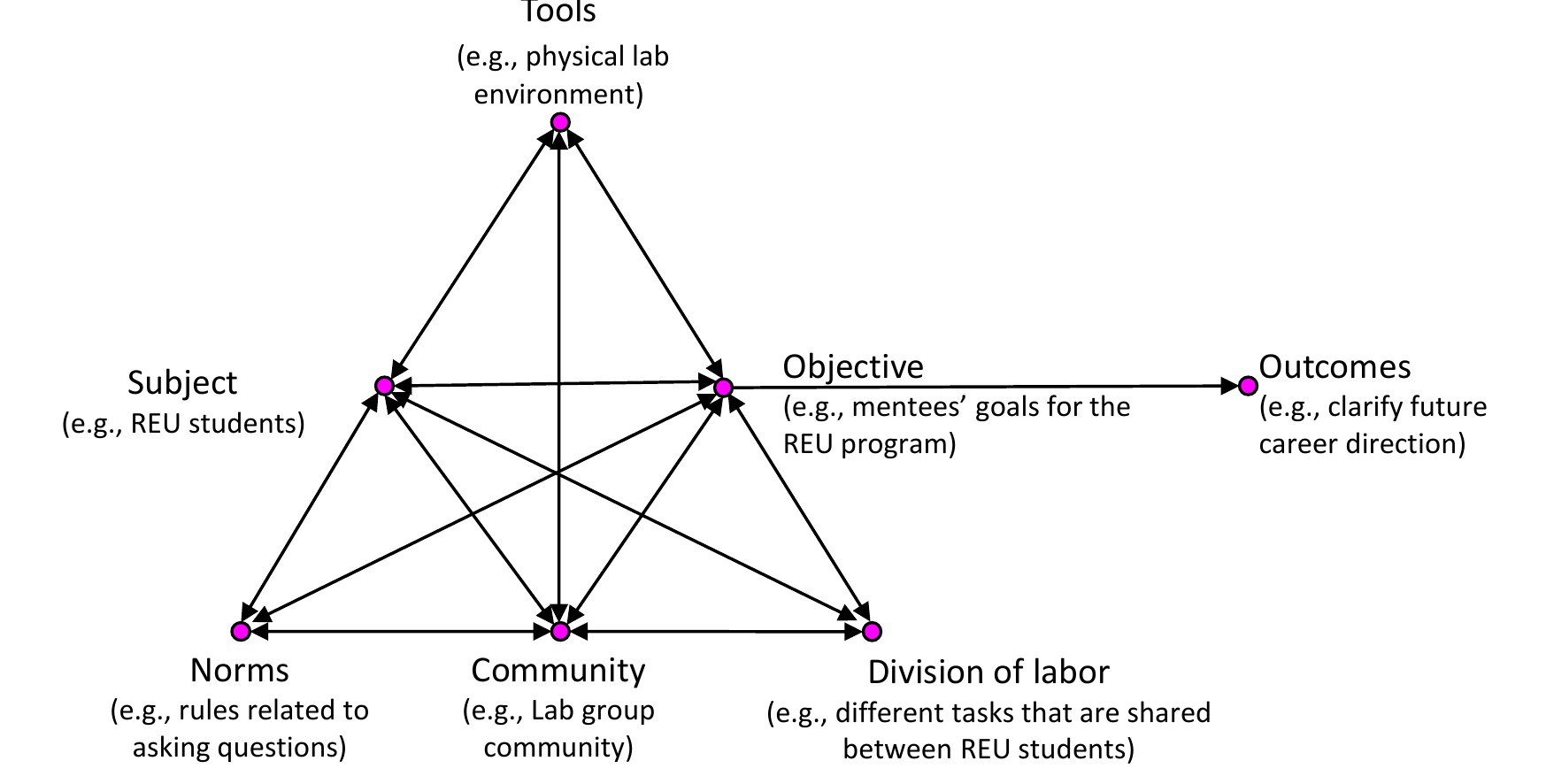}
\centering
\caption{An summary of an activity system is represented as a triangular model \cite{engestrom_learning_1987}. The Research Experiences for Undergraduates (REU) program is an activity system. \textit{Subject} refers to mentees who participated in the remote REU programs. The \textit{tools} include lab environment, software, and REU activities such as seminars. The \textit{objective} is a goal for the remote REU programs. The \textit{norms} refer to embedded rules among members of the community. The \textit{division of labor} is how tasks are shared between members of the lab group. The \textit{community} refers to all members of the research lab and REU participants. The \textit{outcome} of the REU programs is the result of participation in this remote REU experience.}
\label{ActivityTriangle}
\end{figure}

Despite differences in the individual goals of each research project and research program, there are several common goals between all undergraduate research programs; they all hope to increase retention in a STEM career pathway, promote STEM knowledge and practices, and integrate students into STEM culture \cite{national_academies_of_sciences_undergraduate_2017}. While many studies show these goals are often accomplished through UREs, they do not fully describe why a URE would lead to increased retention in STEM among participants. In most studies of URE outcomes, data have been derived from self-reported surveys \cite{hathaway_relationship_2002, blockus_strengthening_2016}, and end-of-program formal evaluations \cite{hathaway_relationship_2002, blockus_strengthening_2016, lopatto_undergraduate_2007, russell_benefits_2007}, while fewer studies have used in-depth multiple interviews \cite{seymour_establishing_2004, hunter_becoming_2007}. 
Furthermore, most studies have exclusively explored the outcomes of in-person undergraduate research experience, and very little research has yet focused on a remote research experience \cite{forrester_how_2021}. We designed the current study to identify the range of challenges that students experienced and characterize some of the possible outcomes of the remote REU program. We used longitudinal semi-structured interviews with mentor-mentee pairs over the undergraduate summer research programs. In the current study, we used Cultural-Historical Activity Theory (CHAT) \cite{leontev_problem_1974, roth_vygotskys_2007, engestrom_learning_1987} analysis to describe the challenges and outcomes of the remote research experiences. The Research Experience for Undergraduates (REU) program as a whole unit it modeled as an activity system. Fig.~\ref{ActivityTriangle} shows a simplified representation of the REU as an activity system in the standard CHAT format developed by Engestr\"{o}m, which represents the system as a triangle that connects a subject to their objectives and outcomes \cite{engestrom_learning_1987}. %The mentee is the subject of the REU activity system with the ability to perform actions and achieve their objectives. Activities are directed toward objects, which include mentees' goals for their REU projects in addition to their personal goals. 
The research questions that guided our analysis were:\\
\begin{itemize}
     \item RQ1: What challenges were observed within the remote goal-directed REU activity?
     \item RQ2: What are some of the outcomes of the remote REU programs?
\end{itemize}
%trim={<left> <lower> <right> <upper>}
% how we can view challenges as potentially resulting from contradictions or tensions within the activity system

\section{Theoretical background}
\label{Sec: Theoretical background}
Our theoretical framework applies Cultural-Historical Activity Theory (CHAT) \cite{leontev_problem_1974, roth_vygotskys_2007, engestrom_learning_1987} to help us reveal the challenges and obstacles in the remote REU setting, as well as identify and describe the possible outcomes of research practices in a remote environment. Engestr\"{o}m's third-generation Cultural-Historical Activity Theory (CHAT) argued that every human activity is goal-oriented and mediated by different components such as tools, division of labor, rules, and community (see Fig.~\ref{ActivityTriangle}).

The links between the elements indicate that they are dynamic and interact with the other components to describe how students' learning can be experienced across diverse educational settings \cite{engestrom_learning_1987}. Subject refers to an individual or group whose perspective is considered for analysis. For instance, we identify each individual mentee as a subject with particular personal inputs (e.g., emotions, family background). The objective is the goal that is the focus of the activity system. In our study, objectives are mentees' goals for their REU projects and participation in undergraduate research programs. Tools are material, symbolic, and conceptual resources used to meditate between the student and their goals. For example, tools include the physical lab environment, mathematics, software, professional development seminars, and experimental kits. The community refers to the people who have shared goals, including research mentors, lab group members, REU participants, and possibly a larger scientific community. Norms refer to the values, expectations, and guidelines for the subject to participate effectively as a community member. An example of a norm within the research group might be how often and by what means the mentee should reach out to the mentor or other group members to receive help. The division of labor describes the different roles performed by the community as they work toward a common objective. For an REU, the division of labor is how different tasks are shared between REU students and their mentors or other lab mates. Lastly, the outcome refers to the results of participating in an activity such as experiencing the remote REU programs. 

One useful feature of a systems-level analysis is that it can reveal aspects of the system that have conflicting influences on the activity. The observable effects of this misalignment are referred to as tensions that result from an underlying systemic contradiction. A plausible example would be a mentee who wants to be recognized by their research mentor for doing a good job at their research. If the mentee believes that asking questions is a sign of ignorance (a norm) then they may avoid asking questions to make it appear they understand more and gain their mentor's approval. However, by not asking questions and communicating with their mentor, they may be less likely to progress in their scientific research and have a successful mentor-mentee relationship, which ultimately lessens the chance they will receive the recognition they desire for their accomplishments. Hence, there is a contradiction between the norms (question-asking indicates ignorance) and their use of the community (mentor) to help accomplish their goals (recognition). 
Generally, tensions, problems, or frustrations that affect the achievement of goals can be interpreted as contradictions within and between nodes in the activity system. Identifying contradictions is important because it provides a mechanism that explains why the problem occurred, which can be used to improve remote REU activities in the future. %We interpreted challenges as a contradictions between two linked components of the activity system that could affect the goals of the REU activity. %Fig.~\ref{Fig_2} shows how our two research questions are situated within CHAT. This framework allowed us to understand the challenges and obstacles that mentees faced in the REU activity due to the tension between any of the component of REU activity system (see the red lightning bolts in Fig.~\ref{Fig_2}(a)). 
%These other components of the activity system are depicted as nodes of intersection in a triangular figure. We could then simply discuss the REU experiences as an activity system that leads to commonly desired outcomes outcomes across many UREs (see Fig.~\ref{Fig_2}(b)).

%\caption{\textit{Subject} refers to mentees who participated in the remote REU programs. The \textit{objective} is a goal for the remote REU programs. The \textit{outcome} is the result that mentee are expected to achieve. The ``tension'' icon denotes the impact of contradictions between two linked components of the activity system that could affect the outcomes of the REU activity.}
This study is a part of an in-depth examination of an REU from multiple perspectives including psychosocial growth \cite{zohrabi_alaee_impact_2022}, challenges and outcomes (this paper), and a more comprehensive description of the tools, communities, norms, and division of labor that occurred with remote REU programs (forthcoming). This paper focuses on findings around challenges and outcomes of the remote research experiences from multiple interviews with undergraduate physics mentees and their mentors who participated in a remote REU program in the summer of 2020. %We discuss outcomes from remote REU without focusing on the mechanism. %of why those outcomes are shaped. 
%We focus primarily on some of the challenges and barriers of doing research in a remote setting  and the outcomes that were shaped in such a remote environment.

\section{Methodology}
In this article, we explore REU programs from both mentees' and their paired mentors' perspectives. Both mentors and mentees were aware of each other's participation in this study. The interviewer assured students and mentors that their own perspectives would be confidential and not divulged. We adopted a qualitative longitudinal phenomenographic approach \cite{Marton1997, marton2000structure} combined with the CHAT framework \cite{engestrom_learning_1987} to collect data throughout the REU program and after it finished. The CHAT lens help us to organize and interpret relationships between multiple components of the REU activity system. The phenomenographic aspect of our study design aimed to investigate the variation in the ways that the mentees talked about each components of the remote REU activity in addition to the outcomes. The longitudinal design helped us to capture many features of the activity system and examined students academic and psychosocial growth over the time of the program, which was an aim of our extensive analysis \cite{zohrabi_alaee_impact_2022, Zohrabi_perc2021}.

%In phenomenography the purpose is to describe the qualitative variation in people's experience \cite{Marton1997, marton2000structure}. 

\subsection{Data collection}
\begin{figure*}[t]
\includegraphics[trim=5 70 10 20,clip,width=180 mm]{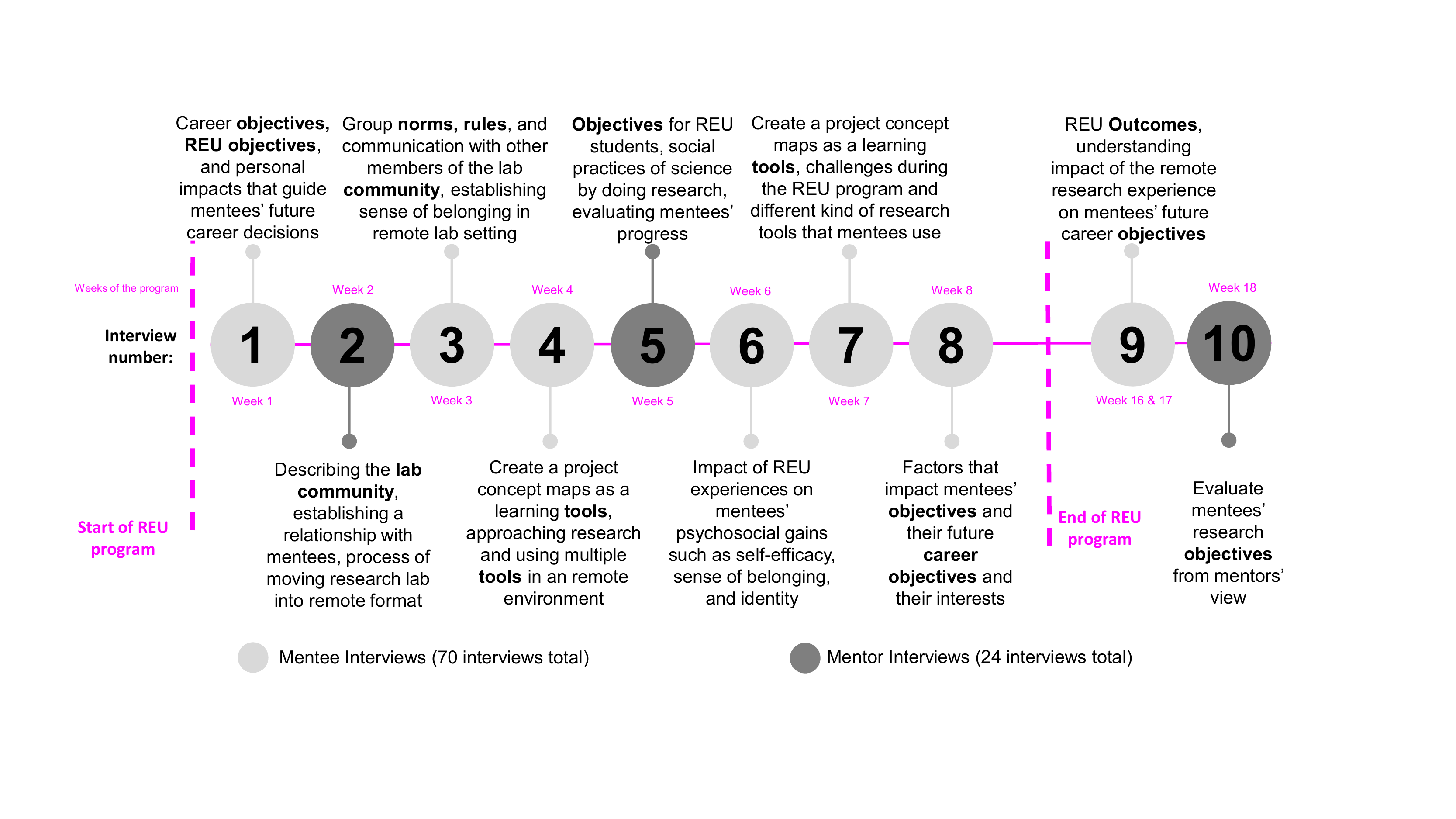}\caption{Timeline of longitudinal study that explored multiple facets of the REU as an activity system. The bold words indicate intentional alignment between the interview protocol and the CHAT framework. The interviews included perspectives from mentees (light gray) and mentors (dark gray). Interviews 9 and 10 occurred after the REU was complete and emphasized many of the outcomes that are discussed in this paper. } \label{ProjectTimeline}
\end{figure*}
%{<left> <lower> <right> <upper>}
We sent 64 physics REU program coordinators an email asking if their REU would be offered in a remote format in the summer of 2020. We received eight positive answers. Mentees were recruited into this study by an invitation email from their REU coordinators on behalf of us. After students volunteered to participate in our study, we contacted their mentors. Our overall sample included ten mentees and eight paired mentors from six REU programs. Demographics of all participants are shown in Table~\ref{tab:demographics-participants}. The sample of mentees was gender and ethnically mixed, while the sample of mentors was all men. We were unable to recruit women mentors. %One mentee had a woman mentor, however the mentor declined to participate. 
While the percentage of women among physics and astronomy faculty members has grown in recent years, in 2019, 19\% of physics faculty members and 23\% of astronomy faculty members were women \cite{AIP2019}. Since the COVID-19 pandemic started, all faculty were negatively affected \cite{AIP2022}. However, women faculty often faced greater challenges due to additional responsibilities of caregiving for other family members \cite{myers2020unequal, katz2021re}. All mentees were physics majors. Additional information about the mentees' projects can be found in the Appendix section (all participant names are pseudonyms). The students were compensated with a \$20 gift card for their participation in each separate interview, which was sent weekly.

\begin{table}[thb]
\begin{tabular}{l|l|c}
\hline 
\hline 
\textbf{\makecell[l]{Mentees' \\Characteristics}} &{}& \textbf{\makecell[c]{Number\\($N$=10)}}\\
\hline 

 \makecell[l]{Gender}	& \makecell[l]{Women}	& {4} \\
{}	& \makecell[l]{Men}	& {6} \\ 
\hline 
	& \makecell[l]{White}	& {5} \\

 \makecell[l]{Race}	& \makecell[l]{Asian}	& {3}\\
{}	& \makecell[l]{Mixed}	& {2} \\\hline
 	& \makecell[l]{Rising senior}	& {7} \\ 
\makecell[l]{Year of college}	& \makecell[l]{Rising junior}	& {2} \\ 
		& \makecell[l]{Rising sophomore}	& {1} \\ \hline 

 \makecell[l]{Type of home}	& \makecell[l]{Ph.D. granting institutions}	& {4} \\
 \makecell[l]{institutions}	& \makecell[l]{Master's granting institutions}	& {2} \\
   \makecell[l]{}& \makecell[l]{Bachelor's granting institutions}	& {4} \\
\hline\hline 
\textbf{\makecell[l]{Mentors' \\Characteristics}} &{}& \textbf{\makecell[c]{Number\\($N$=8)}}\\
\hline 
 \makecell[l]{Gender}
	& \makecell[l]{Men}	& {8} \\ 

\hline 
 \makecell[l]{Type of REU}	& \makecell[l]{Doctoral Universities}	& {6} \\
\makecell[l]{institution}
		& \makecell[l]{Baccalaureate Colleges}	& {2} \\
\hline \hline 
\end{tabular}
\caption{Participants' characteristics}
\label{tab:demographics-participants}
\end{table}
All participants were individually interviewed at multiple points throughout the REU program in the summer of 2020 and one time after the REU program finished (interview nine with mentees and interview ten with mentors). We video-recorded all interviews via Zoom with the permission of the interviewees. Mentees' interviews took between 60 and 90 minutes, while mentors' interviews took between 30 and 45 minutes. Overall, 94 interviews were conducted. Fig.~\ref{ProjectTimeline} shows the overall protocol content for each week of the interview and how it was aligned with the CHAT framework. 

\subsection{Data analysis} 
Each interview was recorded and auto-transcribed with Zoom for analysis. After the interviews were completed, the transcripts were cleaned to fix errors and punctuation. The transcripts became the focus of our phenomenographic analysis. To answer the research questions posed in this article, we focused on the REU activity system as a whole to understand mentees' desired and observed outcomes and their challenges. 
\vspace{-5mm}
\subparagraph{Analysis needed for research question 1.} 
To address research question 1, we needed to identify challenges that mentees experienced during the REU program. Mentees were asked directly about challenges in multiple interviews. However, many of these challenges also came up when mentees reflected on their research experiences more broadly. The language that mentees used to describe challenges included emotive difficulties (e.g., ``it's kind of frustrating''), procedural difficulties(e.g., ``when they're not working right, I guess it has been one of the big hiccups''), and social difficulties (e.g., ``it's hard to get feedback''). 

%Challenges were disruptions in the REU activity system, so identifying them may provide opportunities for developing an improved version of an online REU. 
The analysis process is divided into multiple steps, including data immersion, identifying CHAT nodes and coding them, searching for sub-codes, and reconstructing an activity system diagram to understand the challenge. The process of data immersion started with transcribing, cleaning, and listening to the interviews repeatedly. Analysis of the transcripts was executed using Dedoose software \cite{noauthor_dedoose_2018}. In the primary analysis for the RQ1, DZ identified contradictions based on what mentees explained to her when they were facing challenges or other difficult situations. After several rounds of discussion between DZ and co-another BMZ to reflect on the challenges, we noticed that tensions could occur both within one element of the activity system, such as a lack of community, or between components of the activity system, such as a lack of connection between tools and objectives. We used the CHAT lens to reconstruct the features that led to the particular challenge. CHAT helped us to examine the interactions between the nodes that participants mentioned during their research experience. 

\subparagraph{Analysis needed for Research question 2.}
\label{Sec:question 2}
Mentees were asked directly about the outcomes of the REU experience at successive points during the REU program and after it finished. Most of the results from this research question emerged from interview 9 and 10 which focused on the impact of REU experience on mentees' academic goals from mentors' and mentees' perspectives. To address research question 2, DZ identified moments when participants talked about the outcomes of the REU program based on what they achieved and how they thought this remote research experience would impact their professional and academic decisions. DZ then applied a phenomenographic analysis process by reviewing every excerpt under the outcome node and searching for variation and significant themes. Next, DZ sorted them into themes based on similarities. Our main goal when using phenomenographic methods was to gain insights into our participants' perspectives regarding desirable and possible REU outcomes. DZ met regularly with BMZ to discuss the coding process.
Finally, in the discussion (Sec.~\ref{sec: Discussion}) we compare our results elucidating the outcomes of remote research experiences with previous findings on the outcomes of in-person research experiences.

%This initial analysis identified a collection of quotes coded under the outcome node. 
\section{Results on research question 1} 
\label{Sec: Challenges rq1}
\textit{\textbf{What challenges were observed within the remote goal-directed REU activity?}}\\
The COVID-19 pandemic impacted both mentors and mentees who participated in the remote REU program. They all joined this program with their own expectations, concerns, and emotions about the new remote research experience. In general, challenges included factors that may be due to unique circumstances around COVID-19, particular features of the remote format, or could occur in any research opportunity. For instance, Grace said, ``I know we were very delayed and getting our access to the supercomputers at the beginning, which was a little frustrating''--Interview 9, (Solid State Physics). This challenge was not only due to the remote REU format, since the supercomputer access is typically remote. Additionally, COVID-19 likely have disrupted other workers who were responsible in helping obtain supercomputer access. %We give several examples of ways in which the remote format caused extra challenges for mentees.
This section will provide several examples of ways in which the remote format both created unique challenges for students and exacerbated existing difficulties for REU students.
\begin{figure*}[t]
\centering
\includegraphics [trim=20 100 100 50 ,clip,width=170mm]{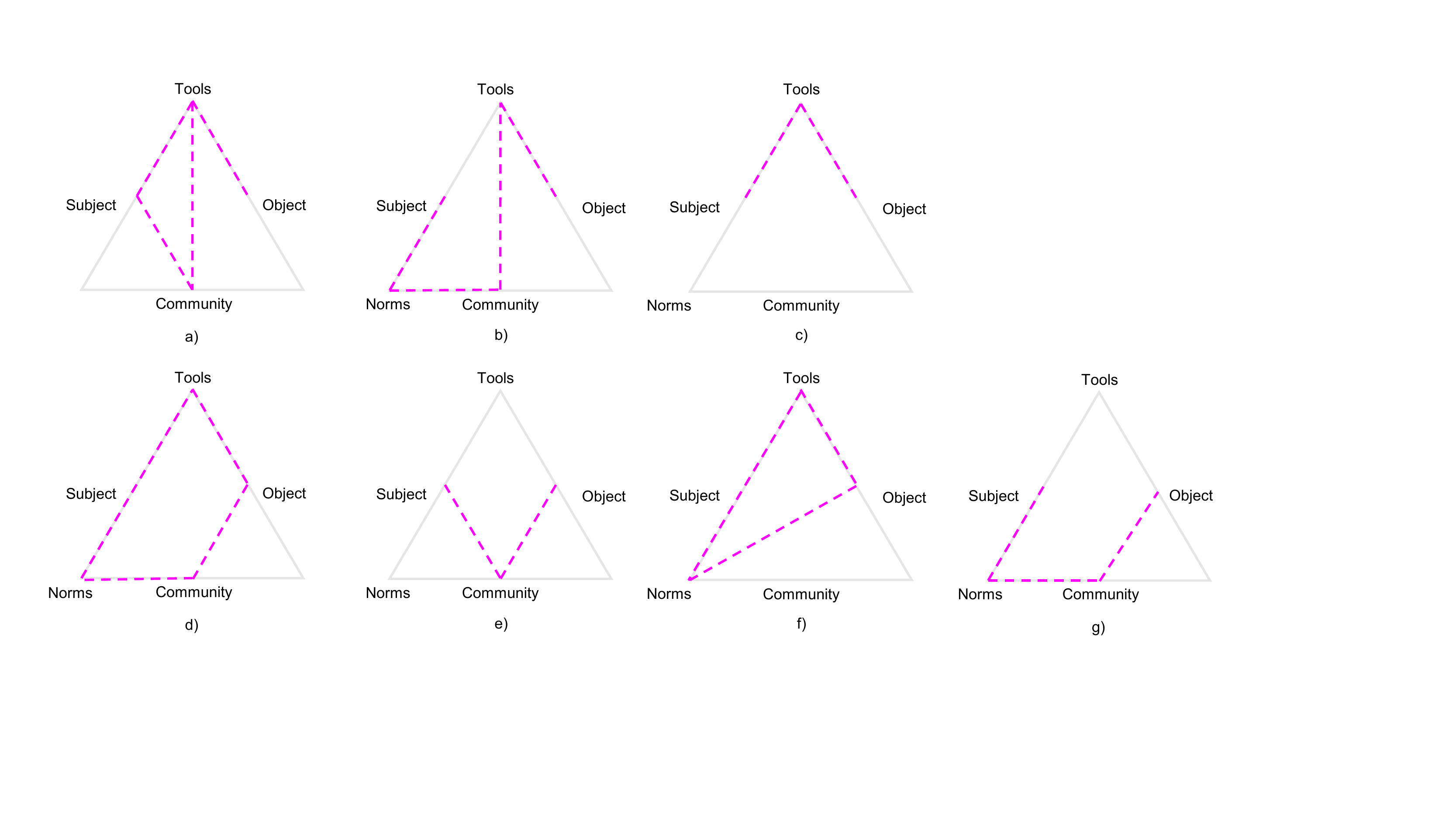}
\centering
\caption{The diagrammatic representation of various tensions between the mentee and their objectives that has been disrupted, resulting in a challenge within the remote REU experience. The dashed lines represent tension between nodes.}
%Dashed lines denotes a pathway between the mentee and their objectives that has been disrupted, resulting in a challenge within the remote REU experience.}
\label{TensionTriangle}
\end{figure*}
\subparagraph{Working from home introduced technical challenges.}
\label{Sec: Facing technical challenges at home}
% Olivia, Brian, Amanda
Several mentees mentioned they had a hard time accessing the right technologies, partly due to the lack of resources and partly due to the lack of in-person contacts who could provide technical support. Fig.~\ref{TensionTriangle}a is a diagrammatic representation of the challenge showing that tools were not able to mediate the research goals due to a lack of access to the proper tools and community support to use those tools. For instance, Helen, who simulated the decay process of short-lived isotopes, had some difficulty trying to ``compile different software... I think the actual processes aren't that difficult, but where I've struggled is like downloading the different software and like trying to become familiar with this new software well. Like analyzing something and I think you kind of understand what the general approach should be, but figuring out how to analyze it with this new software has a learning curve, and downloading it and trying to compile it was challenging... It's hard to do it remotely, I have reached out to my main professor and then also the director because the director helped me set up a remote connection to access the software on his Linux system since it wasn't working on my Mac.''--Interview 4, (Nuclear Physics). According to her, she was struggling with software issues most of the weeks during the REU program; ``I have been frustrated with the whole software thing''--Interview 6. Similarly, by week 3 of the REU program, Grace said, ``We're still waiting for some access to their supercomputers.'' David said, ``One of the most challenging things was getting used to the way the software works because obviously with the modeling software... Especially since I have to do it on my computer as opposed to a more powerful computer.''--Interview 4, (Acoustic). %Due to the importance of technology in a remote research setting, it is necessary to provide more human support and more training than what they received in using technology effectively and fulfilling their project goals.

\subparagraph{Working from home introduced motivational challenges.}
\label{Sec: Working from home introduced motivational challenges}
The sudden transition from working in a lab environment to working at home in pajama pants and hanging out with family members introduced several motivational challenges. A lack of social interaction (working from home in isolation) and lack of communication with other members of their research lab also made some mentees feel discouraged. Fig.~\ref{TensionTriangle}b represents that apart from the remote workspace issues, the lack of work norms and communication reduces the level of motivation and impacts mentees' progress towards achieving their goals. Mentees were usually asked by their mentors to embody the work norms of a research lab (e.g., staying on task, reaching out). However, it was harder in such a remote format without seeing other members of their lab embody those same norms, or feeling some of the ``peer pressure'' to work. 
For instance, Helen said, ``In terms of seeing the end goal, I haven't really necessarily seen that. Sometimes that's been hard since I feel like I need to know what I'm doing is on the right track. But I think also I'm doing less than what I want to do. That's like working online. It's kind of hard to know what point, you should be working on. Like, what other people are up to in their programs as much because you don't get to see them every day and have the subtle interaction'' --Interview 6, (Nuclear Physics). 
%Helen stated that doing research in the remote setting was hard, ``I's harder when you're not like around everyone like doing the research…It's kind of harder because different things are happening in your house or like in the world.  You just don't have that same motivation sometimes from your environment.’’ In order to deal with those down days and those feelings, Helen said, ``I attended the workshop from American Physical Society about emotional resiliency and stuff during COVID-19 and research. That was helpful to hear from other like physics researchers and students who are doing like research over the summer remotely’’ --Interview 9, (Nuclear Physics).  

A lack of social interaction made Bruce feel more isolated, which reduced his productivity. He said, ``Being personally motivated was kind of difficult... Because I was living at home. There are currently five people at home. Most of the rooms are taken up by other people's activities... There were certainly some things that were distractions. Now that I'm looking back when I wished I had gone outside more. Because I really kind of got clumped inside and that probably wasn't great for me. I felt like my mind got stuck sometimes because of that. My parents and my sisters were constantly trying to ask for my time as well as trying to convince me to go out and stuff like that. I lost a lot of time in that sense as well''--Interview 3 \& 9, (Quantum Nonlinear Optics). Fig.~\ref{TensionTriangle}c shows that how the lack of the physical and social lab environment of Bruce impacted his motivation. 
%Working remotely brings its own set of norms and challenges that may impact students' productivity. Lack of motivation may be intrinsic to a remote research experience, but it's important to identify these challenges in order to mitigate future problems and increase self-motivation. Mentors and REU coordinators need to ensure that mentees have all the physical and social tools they need (e.g., good chair, comfortable work area, and the best communication technology to facilitate communication) to develop a strategy for moving forward and handling the isolation challenges. In addition, mentors and REU coordinators can plan social activities outside of their research projects such as a movie night. 

\subparagraph{Information overload and lack of time introduced challenges.}
\label{Sec: Information overload and lack of time}
Even the most motivated and organized mentees can get frustrated by the amount of new essential knowledge that they have to learn in a limited time of the REU program. As one example, Joshua had experienced frustration with his lack of preparation. He said, ``The reason is that I need to learn a lot of basic background knowledge. So, that is hard''--Interview 9, (Nuclear Physics). Similarly, Bruce said, ``I am pretty limited on time... I know there's a lot more information to go on it, but I just haven't really done that and dissuaded from moving in that direction, just because of the limited time I have''--Interview 7, (Quantum Nonlinear Optics). The mentees' REU experience can have competing goals such as learning more physics knowledge and completing the research project goals. Frustration can arise when mentees are trying to gain a deeper understanding of their project content from one side and still need to finish their project at the end of the program from the other side. 

While it can be difficult for mentors to determine when mentees feel frustration during in-person research programs, it is even harder to understand these issues in a remote format when mentees are not physically present in a lab. Mentors don't see mentees most of the time and they can assume how are mentees doing based on their body language, and facial expressions in Zoom meetings, or directly ask them about their challenges. The second issue is the missing learning opportunities from not being physically present in an environment all focused on a topic (the books, the chats, and the whiteboards all provide learning opportunities in less formal ways). This informal communication allows them to learn a large number of background materials. %To complete the assigned project in a timely manner mentees need to identify the required knowledge and develop a project timeline. 
%Bryce w3 t's a huge like  There's a very big jump. You had to make in that first like one to two weeks of just getting versed in the background knowledge of what we're working on. Because, I mean,  None of that like you cover the basics of it in classes that college and science classes I've taken previously, but I mean the specifics of it and how in depth that goes, You don't come anywhere close to touching. So, I mean, the  The gap that I had to close was bigger than I expected.

A lack of understanding can make a mentee feel like an outsider in the lab. Namely, a lack of enough background knowledge on the project topic may lead mentees to experience a poor sense of belonging to the research community. Fig.~\ref{TensionTriangle}d shows that broken links between the top portion of the triangle combined with the lower portion of the triangle (e.g., sense of belonging to the community). Norms also play a role given the mentee and mentor might have differing expectation about the appropriate balance of learning content and accomplishing research goals. Frieda talked about the efforts her mentor made to identify and teach the essential knowledge, which supported her sense of belonging.

For instance, Frieda talked about her mentor who asked her to join their big research group meetings with all the other researchers in the field. She recalled her mentor telling her that, ``You're going to come to these group meetings and you're not going to understand most of what they say, but that's okay. We're going to get there.'' She continued ``Mostly I've been listening because, I don't know that much yet... I think that he just preparing me, so hopefully I will [know] eventually and be able to contribute as well. After each of those meetings, my mentor has met with another REU student and me and worked through some of the terminology and stuff [that they talked about in a big group meeting]. He's doing a really good job making us feel like we can understand it''--Interview 3, (High energy physics). Although this challenge is likely an issue in every REU program, due to the remote setting and lack of face-to-face interactions, mentees are likely to feel less connected and less cared for or accepted by their lab group. 

%mentors and REU coordinators can develop a mentoring strategy for helping their students to move forward in their research experience. 
\subparagraph{Lack of communication opportunities among REU students introduced challenges.}
\label{Sec: Lack of communication opportunities among REU students}
All REU programs have the common goal of preparing undergraduate students for careers in STEM fields by integrating them into research projects and building scientific community of peers. Different REU programs may develop their own plans and structures to achieve these goals. However, our data indicate that REU participants persistently lacked community throughout the REU program. 

Sometimes tensions in the REU system happened. For example, a particular REU program or the REU activity may aim to pursue multiple goals at the same time, such as introducing students to different research areas (e.g., professional development activities) and creating a community of peers (e.g., opportunity for all REU participants to get together). Fig.~\ref{TensionTriangle}e shows in addition to developing a community among REU students, the overarching goal of many undergraduate programs is to provide a long-term benefit by presenting professional development activities. Although, the professional development activity brings everyone into the same Zoom room, it doesn't really form community among students. According to Bruce, ``[Having a community] really has kind of fallen by the wayside, to the point where there has not been a single chat in the group chat since the first week.'' He continued that during social events and seminars, the REU students did not interact because they were listening to lectures, and he wished they had some sort of `all groups' meeting'' where they could present their work to each other. Similarly, Frieda said, ``We have bi-weekly meetings on Tuesdays and Thursdays. At first, there was one about grad schools. And then, we each gave five-minute presentations about our research. Then there was one about each area of research that's going on at the REU institution. So they got a representative professor to come talk for the meeting''--Interview 3, (High Energy Physics). Although the REU coordinators aimed to provide career support during these group sessions, these sessions sometimes had more of a focus on information delivery rather than peer discussions and interactions between mentees. 
\subparagraph{Barriers in mentor-mentee relationships introduced challenges.}
\label{Sec: Barriers in mentor-mentee relationships}
A good mentor-mentee relationship is key to having a successful REU experience. Analysis of the data identified that most mentees maintained a positive relationship with their mentors. However, due to a lack of tools and lack of physical proximity in a remote lab setting, it was harder to form this good mentor-mentee relationship.
%zoom issues
Over the course of the COVID-19 pandemic, meeting over instant communication apps such as Zoom became the norm for communicating remotely. However, these tools have limitations and shortcomings. For example, the Zoom platform was blocked or had a bad connection in China. Joshua's mentor said, ``We use Zoom. But, we also use WeChat. Because sometimes Zoom does not work very well, we lost connection''--Interview 10. 
%not a priority for a mentor

Having multiple weekly meetings was one of the productive norms among many REU groups in a remote lab setting. Weekly check-ins were vital to support mentees' research goals and helping them develop more open communication with their mentors. However, for some mentees, they felt they were not a priority for their mentors and the way they communicated with their mentors was mostly via email. For instance, Helen said, ``We zoomed in the beginning, and I've talked about what my role would be. Then we've just been emailing out the next plan for what I'm doing. I know he has two kids. I think that it's just been like a little hard for him to stay in touch as much''--Interview 3, (Nuclear Physics). This challenge was probably unique to COVID-19 due to the juggling of childcare, school, and full-time work responsibilities with children at home. Fig.~\ref{TensionTriangle}f shows the lack of communication tools and physical proximity can impact both mentor's and mentee's contribution to the new norm of weekly check-ins in the online setting.

Ivan also didn't have any weekly meetings over Zoom, but for a different reason. Ivan lived in a very different time zone from his mentor. Ivan said, ``We do not have virtual meetings. We send emails, and my mentor sends emails back''--Interview 3, (Nuclear Physics). Ivan’s mentor described challenges when collaborating with him, saying, ``I should have told the REU director that I couldn't manage a student who is in a 14 hour different time zone. Unfortunately, the only available hours for me to meet with Ivan overlapped with my meeting hours with people at CERN. In the past, I had always managed the REU time commitments by meeting with people in CERN in the morning and the REU students in the afternoon. Because they were local and that was just not an option [this summer] and there just weren't enough hours in the day for me to meet with Ivan''--Interview 10. %It is important for both mentors and mentees to contribute to the new norm by keeping communication ways open. 

\subparagraph{Differing expectations between mentors and mentees introduced motivational challenges.}
\label{Sec: Differing expectations between mentors and mentees}
Barriers also arose when the expectations of mentees appeared to be different from mentors' expectations. Ivan, who was also enrolled in summer courses at his home institution, said, ``I had to take lots of courses and I did not have too much time to do research... I was too busy to do that... My mentor gives me lots of information that I can read it on my own. I have to use lots of time to do it. He treats me as a researcher and a physicist and he gave me lots of raw materials and I have to do it on my own''--Interview 6, (Nuclear Physics). 

On the other hand, his mentor thought ``It was hard to really get his attention and get him working on the projects as a remote student with the time zone difference and apparently a miscommunication about the amount of time that he was supposed to have budgeted for the REU experience... I don't really know that he has a good idea of what it's like to be a researcher and to work in high energy physics because the validity to pursue a research question fairly independently is a pretty central aspect to that experience and I don't think he got there''--Interview 10. The representation for this challenge is very similar to Fig.~\ref{TensionTriangle}b that shows that program coordinators, mentors, and mentees need to communicate over norms and rules in the research labs in order properly allocate time and gain the major goal which is having a positive research experience and finish the research project. For Ivan, he stated, ``I am just a follower in this program''--Interview 9, (Nuclear Physics). However, his mentor expected him to do some more independent work and dedicated more time to his research.
%serious vs. formal
%scale of the project
Interestingly, differences in expectations between Ivan and his mentor were about more than how time was used on the project. Ivan thought his research group was not serious and formal, which perhaps affected the time and effort he dedicated to his REU. He said his lab members were ``Very kind and funny, [but] it is just different from my expectations''--Interview 3, (Nuclear Physics). He said, ``My mentor gives me some tasks or some problem to solve and some information about those tasks... My mentor just sent me some instructions and if I am working along those directions, I get the work done''--Interview 3 \& 6. He wished his mentor asked him to bring ``some ideas about doing some research.''. His feeling about his project did not change over the REU programs. After the REU program finished, he said, ``I just follow some things just have been done before. So maybe not very big [research], it is maybe just a little part in this size''--Interview 9, (Nuclear Physics). Meanwhile his mentor thought ``He might have been a little bit disappointed with the research project he had because I think he wanted to look at big overarching questions, and we wanted him to look at the data that came out and some checks... We all have dreams of sitting in a room and dreaming up some theory that explains everything, and there is nothing tedious about that process. However, when you actually get to work on an experiment, that is tedious, and it is required to make progress. You have to learn a certain amount of discipline and commitment to seeing through the tedious part so you can get to the fun parts of having a discovery or learning something''--Interview 10. The representation for this tension in Fig.~\ref{TensionTriangle}g shows that both the mentor and mentee did not speak regarding their expectations and did not communicate properly as a group to consider each other's goals for this REU program. For Ivan, he expected to do a bigger project and test his ideas in the project other than following some instructions and solving some problems. On the other hand, his mentor expected him to check some codes. 

The mismatched expectations and norms between mentors and mentees were not a very common theme across our data set. However, the authors mentioned this challenge here since the mentor-mentee's communication is extremely important factor in the students' REU experience and their decisions around their future career. 

%\subparagraph{Tension between the structure and helping students to become an independent researcher}
%It is important to mention something is hard, that alone is not sufficient to be coded as a challenge. For example, Bruce said, ``research definitely is more frustrating, but it's also more rewarding. Things definitely go off the rails, a lot more often. But, it's also more fun because of that. Maybe fun is not the right term to use. It's more exciting. It feels like that your accomplishments are a lot more by your own''--Interview 9. 
%This example shows a tension in project-based work where the outcome is less structured, more opportunity for unanticipated difficulties (going off the rail), but also more opportunity to be exciting and feel a sense of ownership.
%add bryce

\section{Results on research question 2} 
\label{Sec: outcomes rq2}
\begin{figure*}[t]
\includegraphics[trim=10 250 10 70,clip,width=180 mm]{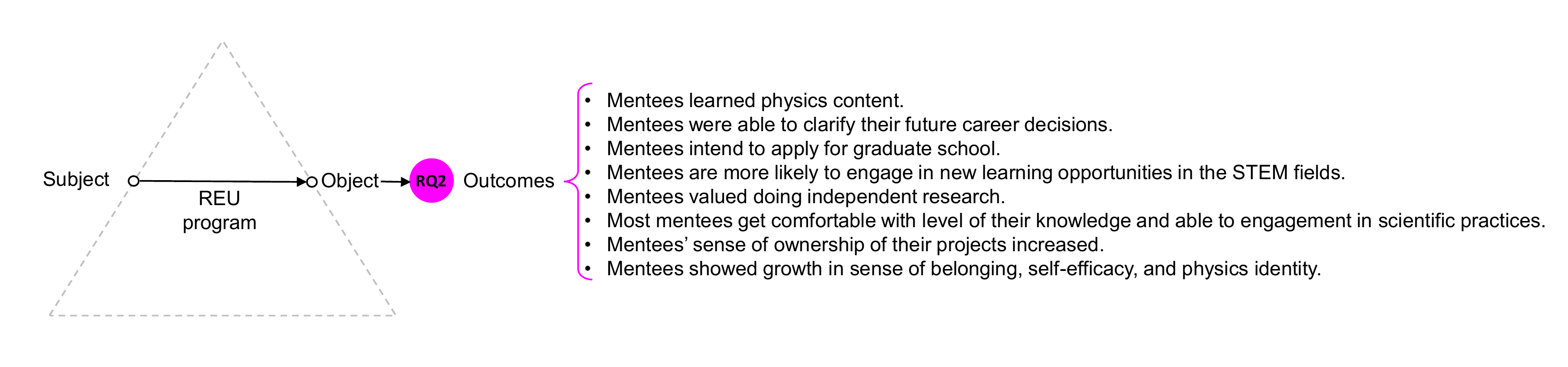}
\caption{Overview of the outcomes of the remote REU programs.} 
\label{REUOutcomes}
\end{figure*}

\textbf{\textit{What are some of the outcomes of remote REU programs?}}\\
REU programs are multifaceted learning activities that involve challenges and benefits. Outcomes are the final part of the goal-oriented REU activity system. A few weeks after the program finished, in the 9$^\textrm{th}$ interview, we asked all mentees to describe the outcomes of their REU experience. We found that almost every mentee, despite their different projects and different circumstances during COVID-19, described having achieved their goals as a result of their research experience. They all stated that attending the REU program was the right decision in their academic journey. The following subsections explain several of the specific outcomes of the remote research experience (see Fig.~\ref{REUOutcomes}).
\subparagraph{Mentees described learning physics content.}
All mentees expressed this remote research experience deepened their knowledge in discipline-specific project content and also introduced them to new areas. An example is Grace, who said, ``I am enjoying it a lot. I am learning a lot about chemistry and biology and just how much physics intersects with everything, and how it is fundamental science. I definitely had some frustrations in some of the earlier weeks reading some of the papers and not understanding them... Feeling like I had to understand everything that was being thrown at me. But, overall, I am definitely enjoying learning these, learning how to use the program, learning the chemistry and physics involved has been really interesting and enlightening''--Interview 7, (Solid State Physics). Joshua, an international student, talked about how much he learned from extracurricular activities during the REU program; ``I learned the skills such as organizing the introduction or other parts of a paper better. In GRE preparation courses, I learned some required physics concepts in English... These are important for my future career. In seminars, I can broaden my horizons, such as I learned some new physics knowledge in other physical areas''--Interview 7, (Nuclear Physics). 

%We also found that getting more content knowledge can impact mentees' self-efficacy. For instance, Helen stated that through remote research during the COVID-19 pandemic, she spent ``more time reading different papers about the same thing. I just feel more confident and know it better... I think I have always enjoyed research, and I enjoy the process. I think undergraduate research is important because I have been exposed to many different areas of physics''--Interview 7, (Nuclear Physics).

\subparagraph{Mentees described their future career decisions.}
The remote REU experience positively impacted most mentees' future career decisions. Most of the mentees ($N$=8) said the REU experience helped them understand the nature of research work and think about their future career in a STEM field. David stated, ``Especially because I was able to get a better idea. Not only I was able to learn more about the software and about the research going on in one of the fields that I am interested in, but also because I was able just to get a better understanding of how research works and that sort of dynamic. I get a better feel for what doing research with a professor looks like. So that I have a better understanding of how that will work when I am, for example, in graduate school doing research''--Interview 9, (Acoustic). Caleb learned that he might prefer a balance of hands-on and computer-based work. He explained he enjoys computational theoretical research, but ``I would like to be in a lab-type setting; some days, I am doing hands-on research, and other days, I am doing computational stuff. My biggest takeaway from the REU program is that I do not think I could just be on a computer for eight hours a day''--Interview 6, (Atomic physics). Due to the lack of information about different career choices and fields at the beginning of the REU program, some mentees were uncertain about they wanted to do. For instance, Ivan said, he is interested in studying nuclear physics in graduate school now, but ``Before the REU program, I did not even know what I was interested in because there were too many fields''--Interview 9, (Nuclear physics).

\subparagraph{Mentees intend to apply for graduate school.}
After finishing the REU program, most mentees ($N$=9) were sure that they would like to apply to graduate programs. Frieda said, ``It was really good for where I was in my education and made me seriously consider this field as a graduate school field, which not that I was not considering it, but I am much more serious about considering it now''--Interview 9, (High energy physics). Similarly, Andrew describes how the REU contributed to desire to do research in graduate school. Andrew said, the REU program ``has influenced me in the fact that it makes me want to do graduate school and research more. Because beforehand, I was like, I do not know if that is going to be a lifestyle I want to get into, because I did not know. I have not done research before, so I did not know if it [would] be something I enjoyed or something I [would] absolutely hate. So after doing this REU, I was kind of like, I enjoyed this. I can see myself doing that''--Interview 9, (Nuclear Physics).

Bruce was unsure about his future career during most weeks of the REU program. However, in the last interview, he said that his mentor provided him with a new perspective and insight on doing research. He said, ``I definitely think that [my REU mentor] has a strong influence on making my decision about graduate school''--Interview 9, (Quantum Nonlinear Optics). Bruce explained that this influence came from stories that his mentor told him about his previous students' career decisions and also from giving him a sense of how he does research. He said, ``There's just so many different ways that they're either the professors or they're working in research labs or in different things... It's just stuff that I really would like to be doing in the future as a career... I think another [influence] is just he's giving me a sense of how he does research... He's not always super stressed about everything. He's not overworking himself. This idea that you can be a researcher and not overwork yourself all the time is also quite enticing''--Interview 9, (Quantum Nonlinear Optics).

\subparagraph{Mentees described engaging in new learning opportunities in the STEM fields.}
Some mentees explained how they planned to tailor their interests toward their future career goals by choosing elective classes, attending the seminars, or reading new related articles and books after the REU finished. One example was Joshua (Nuclear) who decided to take some elective courses in astrophysics and nuclear physics the following semester to narrow down his interests. 

Other mentees tried to read and search through the literature related to their summer research experience. For instance, Helen said, ``I like to read articles. I am part of the American Nuclear Society and get updates and stay informed, especially around medical physics. I always have liked reading Physics World news about medical physics, listening to podcasts, and connecting with people. So I have definitely been more interested in learning more in that area after that REU''--Interview 9, (Nuclear Physics). Similarly, Andrew said, ``What I did over the summer made me want to learn more about that... I got a book on nuclear physics, and I was kind of reading that as the program went along''--Interview 9, (Nuclear Physics).

\subparagraph{Mentees valued doing independent research.}
Due to the lack of in-person interaction between mentees and other members of the lab community, they are had to engage in substantial independent learning. For instance, Caleb said, ``[The remote format] makes [the research] a lot more independent. Because you are still able to reach out via email, Zoom, or whatever platform you use, but there is that added step of composing the email asking for Zoom chat. Whereas in in-person, just like walked out to the office or they are in the room with you, and just like look over your shoulder. So, it adds a lot more independence''--Interview 7, (Atomic Physics). Likewise, David explained that he became more independent in his learning since, ``I will be in the middle of a project, trying to figure something out in the middle of the day. Then I will just go look it up and try to figure it out. I usually use it as a resource on my own, and I think that might be a bit different if I was working with the professor, more closely in a physical environment''--Interview 7, (Acoustic). Although, independence is an important factor in any research work, some mentees were hoping to be more reliant on their mentors, such as Ivan's experience who had a lot of resources from his mentor and wished his mentor let him bring his ideas about doing some research, instead of ``I am just a follower in this program.''
More details describe in section~\ref{Sec: Differing expectations between mentors and mentees}.

\subparagraph{Mentees were comfortable with level of their knowledge and able to engagement in scientific practices.} 
Regardless of the project area, every REU program aims to provide opportunities for students to develop research practices, and become community members by supporting them to work collaboratively. During interview nine after the REU program finished, eight mentees described feeling more comfortable with their ability to get involved in their community (e.g., by asking questions and learning new concepts) and became more comfortable with the level of their knowledge. %Mentees joined the REU with different academic backgrounds and levels of personal recognition. 
%Three mentees stated that because of friendly and supportive lab group dynamics, they became more comfortable with the level of their knowledge. 
For instance, Grace said, ``%A little bit more confident in my capabilities and just to accomplish something,
%Ask question: amanda, kate , adam W9-Q14
%mckenna  yusen, bryce brian and anthony feel more comf in the community
[I can] ask a question and not feel like I am asking a stupid question. It is very hard to look back on my near past self and reflect because... It is hard to see [how much I] change''--Interview 9, (Solid State Physics). David said, ``I had a better understanding of what was going on. So I was able to contribute a lot more''--Interview 9, (Acoustic). Six out of eight mentors also were very happy and satisfy with their mentees' engagement in scientific practices.

\subparagraph{Mentees described increasing in their sense of ownership of their projects.}
In our previously published extensive study, some mentees talked about a sense of ownership as a factor that linked to their sense of belonging \cite{zohrabi_alaee_impact_2022} through scientific contribution. We found that, in order to achieve the ownership construct, mentees need to produce potentially new results for their project. 
Additionally, five mentors described how they intentionally helped mentees' sense of ownership by offering them responsibility in their project. These mentors explained that their mentoring philosophy includes helping mentees increase their sense of ownership. Three mentors mentioned giving their mentees the freedom to challenge themselves intellectually. Caleb's mentor said, ``By respecting and giving them great freedom, that is, say if he had ideas on how to do things. I would say, `Do it and show me.' I did not have to tell them this is how you do it step by step. In the beginning, yes, but it did not take long to get going on his own and then come up with his own improvements and extensions and search. He was very good at that. So I certainly think that encouraged him. I think he liked that. I think he liked the research. But, you would have to ask him, I guess''--Interview 10.

Two other mentors cultivated ownership by providing the bigger picture of the project and giving students responsibilities. As Frieda's mentor explained, ``She had a project, and it was very well specified that this was her project. There was no one else working on it, and again, I am working with graduate students and postdocs, making sure they were not working on the project. So, it was hers and hers alone''--Interview 10. No one else would do what Frieda was working on so the team needed her to get her part done.

\subparagraph{Mentees described growth in their sense of belonging, self-efficacy, and physics identity.} 
A higher level of sense of belonging and identity is associated with improved academic performance and possibly enhanced persistence in the field of physics. Many mentees reported growth in their psychosocial attributes, such as a sense of belonging, self-efficacy, and physics identity. We have already described the psychosocial benefits of these remote REU experiences in more detail in another paper \cite{zohrabi_alaee_impact_2022}. However, for completeness we summarize these benefits here because they were an important outcome of the remote REU experience. We found having a supportive lab research community and mentor-mentee relationship helped most mentees to exhibit psychosocial growth. 

Almost all mentees reported a higher level of sense of belonging to their lab community and possibly to the disciplinary community at the end of the REU program. David, who made a resonator model to learn an acoustic modeling software tool, said he had a ``place in the field now''--Interview 9, (Acoustic). He thought his research experience over the summer positively helped him feel like a part of the physics community. He said, ``Just because I have been able to experience and contribute as well as being able to look through the work that other people have done in the field a lot more in-depth and having done my little bit of work in the field helps me to understand better the work that others have done in it... Even now that I am not doing that research full time, I feel much more in part of that than when I was only taking classes before''--Interview 9, (Acoustic). Frieda said, ``I definitely consider myself part of the physics community... A lot more than I did at the beginning. Because now I feel I know a lot about it, I have gotten a lot done, I am informed, and I feel more in it, like I am more submerged in it personally. Maybe you could say my physics identity has gone from like an amateur to like a beginner to intermediate now. I feel like I am actually part of the field''--Interview 9, (High Energy Physics). Both Helen and Bruce, who had a lower level of sense of belonging during the program, believed they felt more belonging after the program finished since they were back on campus and communicating with more people. For example, Bruce said, ``[I became] more comfortable in the physics community and because of that I became less stressed about interacting with the community... I feel I'm somewhat part of the bigger physics group because I understand how it works now. I am collaborating with some people. I know a finite number of researchers and exactly what they're doing. I've worked with them or I know the kind of research they're doing, so that's why I feel like that.''

Findings for the self-efficacy construct indicate that self-efficacy stemmed from various sources, such as getting more physics content knowledge, doing independent research, producing new results, and scientific communication with other members of the community. For instance, David said, ``I was just trying to learn to catch up and understand what was going on. After I had been doing it for about a month, I felt a lot more confident in what I had been able to learn... I feel a lot more comfortable than before because I had a chance to work full-time doing physics, even if it was just for summer. Since I enjoyed that, I feel a lot more confident proceeding with that as a career goal''--Interview 9, (Acoustic). His confidence stemmed from getting more physics knowledge and contributing to his research group projects.

Sense of belonging to the physics labs or community and physics identity both describe the personal experience around varying degrees of attachment. For example, physics identity can be defined as the self-recognition of being a physics kind of person or a physicist.
%Physics identity is focused on a mentee's perception of self within the field of physics, whether they are seen as a physicist or a ``physics person''. 
For most mentees, developing a more robust physics and researcher identity resulted from a stronger sense of belonging and stronger self-efficacy \cite{zohrabi_alaee_impact_2022}. Positive psychosocial changes that resulted from research also impacted career decisions. During the REU program, 
%Emma, Helen, and Frieda, Ivan, joshua bruce
six mentees believed they were physics apprentices who were on the right path and needed more knowledge and experience to be a physicist. After the REU program ended, five
%W9-Q2yusen, anthony olivia, shuaji and kate
mentees felt a stronger identity around physics and doing research in the field. For instance, Ivan said, ``The REU has taught me something about confidence in doing research. The REU has made you think yourself as a physicist''--Interview 9, (Nuclear Physics). 

\section{Discussion}
\label{sec: Discussion}
%In this study, we first identify some of that challenges of the remote research experience and then transitioned to the benefits and possible outcomes of doing research in such a remote setting.
We had two research questions in this study. The first research question was: \textit{What contradiction were observed within the remote goal-directed REU activity?} We used the combination of phenomenography and CHAT framework to understand the challenges in a new remote complex educational system. Based on our semi-structured interviews, mentees shared with us different types of challenges they faced during the remote REU program. As we outline in section~\ref{Sec: Challenges rq1}, COVID-19 and the remote format impacted students' learning. For instance, essential components of effective learning, such as easy availability of helpful feedback and having a good mentor-mentee relationship, were harder in a remote format. %All mentees experienced a new life in quarantine. Many of them expressed that they had to set up a workspace in different rooms of the house or a basement to stay on task. Thus, these findings are in tandem with the results from previous studies around learning challenges in a remote format \cite{2007AGUFMED33B1226H, gonzalez2021emergency, nambiar2020impact, zhang2020suspending}. In addition, the limited exposure of REU students to the REU community was more challenging. Although each REU program had different social events such as meetings with previous REUs, check-in meetings, group presentations, movie nights, seminars, and elective courses to build community among their REU students, it was still not enough. Most mentees enjoyed the social events, but they would have liked more whole-group interactions. 

\begin{table*}[t]
\begin{tabular}{p{6cm}|p{11cm}} 
\hline 
\textbf{\makecell[l]{Goals expressed in literature for\\ students participating in in-person\\ UREs \cite{national_academies_of_sciences_undergraduate_2017}}} & \textbf{\makecell[l]{Short term outcomes of the remote \\Research Experiences for Undergraduates\\based on our data}}\\
 \hline\hline
{Increase mentee's retention in STEM fields}& {-Clarifying future career interest}\\
    {}& {-Applying for graduate school}\\
     {}& {-Engaging to new learning opportunities in the STEM field}\\

\hline
{Promote STEM disciplinary knowledge}
& {-Learning physics content}\\
    {}& {-Appreciating value of doing research independently}\\
      {}& {-Utilizing disciplinary research practices by asking questions and directing projects}\\
\hline
{Integrate students into STEM culture}
& {-Finding interest in physics field}\\
  {}& {-Expressing higher level of sense of ownership of their project during REU experience}\\
   {}& {-Becoming enculturated in the physics community and expressing a stronger sense of belonging, self-efficacy, and physics identity as a result of the remote REU experience} \cite{zohrabi_alaee_impact_2022}\\
   \hline
\end{tabular}
\caption{Comparison between the primary goals for in-person UREs and the actual outcomes for the remote REU program.}
\label{tab: comparison}
\end{table*}

Our second research question asked: \textit{What are some of the outcomes of the remote REU program during COVID-19?} Although, the first research question identified some challenges, our data still showed that the outcomes of the remote REU program are not too far from the generally accepted benefits of any in-person undergraduate research experience. Table~\ref{tab: comparison} compares the primary goals for UREs described in the literature \cite{national_academies_of_sciences_undergraduate_2017} (left column) with outcomes of the remote research experience from our findings (right column). It should be noted that mentees describe outcomes that are short term (e.g., preparing graduate school application materials) and long-term (e.g., future career decisions). %In the following section, we review the three main URE goals and discuss them individually in light of our data. 
Most previous studies have focused on three major outcomes for in-person UREs: increasing retention and persistence in STEM, promoting STEM disciplinary knowledge and practices, and integrating students into STEM culture \cite{nagda_undergraduate_1998, rodenbusch_early_2016, graham_increasing_2013}. We use these same three categories in Table~\ref{tab: comparison} and link our results (Sec.~\ref{Sec: outcomes rq2}) to each of those categories. %It should be noted that mentees describe outcomes that are short term (e.g., preparing graduate school application materials) and long-term (e.g., future career decisions). In the following section, we review the three main URE goals and discuss them individually in light of our data. 

%Despite the challenges of the remote REU experience, most mentees ($N$=9) gained clearer ideas about their future career goals.
%We noticed that many of the positive outcomes described above in the Sec.~\ref{Sec: outcomes rq2} are similar to the psychosocial and academic growth that previous literature discussed as an outcomes of in-person research experiences.  %In addition to the short-term goals of many URE programs (e.g., improving students' learning gains), the overarching goal of many URE programs is to provide a long-term benefit. 

\subparagraph{Increasing mentees' retention in STEM fields.}
Undergraduate research experiences can impact students' retention in STEM degree programs and provide them with a new way of thinking about their future career paths in STEM fields. In-person UREs often aim to help mentees understand what it means to do research and what a science career might look like \cite{national_academies_of_sciences_undergraduate_2017}. The outcomes associated with this goal across our data included: clarifying future career interests; applying for graduate school; and engaging to new learning opportunities in the STEM field. This result also is consistent with the evidence in substantial prior research that has examined the influence of UREs on students' long term outcomes, such as clarifying future career goals \cite{lopatto_survey_2004, seymour_establishing_2004, hunter_becoming_2007}.

\subparagraph{Promoting STEM disciplinary knowledge and practices.}
UREs can help students to develop new skills, learn new knowledge, and engage in the practices of their STEM discipline, such as using computational models or analyzing and interpreting data. The outcomes associated with this goal in our data include learning content knowledge through the REU experience, gaining independence as a researcher, and developing questioning skills. In terms of the activity system CHAT triangle, the disciplinary knowledge and practices would be considered research \textit{tools} that support the research objectives of a student's project.
Several previous studies on UREs outcomes indicated that meaningful research practices could facilitate students' learning and development of their technical knowledge in addition to promoting their communication skills \cite{junge_promoting_2010, hunter_becoming_2007, kardash_evaluation_2000, national_academies_of_sciences_undergraduate_2017}. Similarly, in our data multiple mentees said they felt like they better understood physics concepts and what was going on in their lab research group. Besides, most mentees were comfortable reaching out to lab members for help. 
\subparagraph{Integrating students into STEM culture.}
Other studies of in-person UREs mentioned that undergraduate research experiences could strengthen students' motivation and interest in STEM culture. Here, integration into STEM can be thought of as a psychosocial growth process (e.g., a gain in physics identity and sense of belonging) where students learn the values of the discipline. Our findings around integrating mentees into STEM culture include: developing interest in particular subfields of physics, a sense of ownership of their project, and growth of psychosocial constructs. 

Integrating into STEM culture requires integration into a community. There is considerable evidence across our data that positive remote REU experiences require community interactions to achieve desired outcomes. Nevertheless, as described in the challenges in Sec.~\ref{Sec: Challenges rq1}, a lack of community and communication with other REU participants persisted throughout the REU program. %Another way of enculturating students in the community is through strengthening interactions between mentors and mentees. Emma, who had previously done an in-person research experience, exemplified how a good mentor-mentee relationship could foster self-efficacy, physics identity, sense of belonging, and ownership. She said, ``I really enjoyed working with my [REU] mentor; I think the dynamic is great. He was my advisor, but he also treated me like an equal, which is uncommon when working with anyone with a doctorate when you are an undergraduate student... My [REU] mentor trusted me to get things done and to have my own ideas. He kind of believed in my ability to do any physics-related thing, which is not super common when people see that I am instructional physics... My [REU] mentor was just like, here is the stuff we are doing, what do you want to do... And he was not like, do you know how to do circuits? Do you know how to use the 3D printer? He was like, `here is the project, let's do it.' And anything that I was unsure of, I would just be like, how do I do this, and he would just accept it and teach me like the 3D printer''--Interview 9, (Physics Education).% I did not know how to use a 3D printer. Now I do... I learned about using 3D printers, which is a lot of fun... He does not question, what you can do, but instead leaves it open-ended for you to come in when you need help'' This emphasizes how mentors can build a solid and trusting relationship with their students. In our extensive study \cite{zohrabi_alaee_impact_2022}, we argued how this communicative relation between mentor and mentee could impact mentees' self-efficacy, sense of belonging, and physics identity, which lead to a positive view on their future career. 

\section{CONCLUSIONS AND RECOMMENDATIONS}
Prior to early 2020, the vast majority of articles about online education debated whether such methods were effective and whether academia should more broadly use remote learning. In a perfect world, maybe! However, when the COVID-19 pandemic began in early 2020, many institutions rapidly experienced a remote learning transition. In this study, We attempted to elucidate some of the challenges and outcomes associated with remote research experiences. Our data shows that all participants described their remote experience as a ``real'' research opportunity that achieved desirable outcomes. The remote format still allowed mentees to meet their personal goals by the end of the program. %Overall, outcomes of the remote research experience were similar to traditional in-person REU experience such as gaining new knowledge and skills and an increased intention to apply to graduate school.

Following here we discuss recommendations to address challenges (see Sec.~(\ref{Sec: Challenges rq1})) and increase the quality of remote research opportunities in the future. 

\textit{Recommendation 1: ``Provide technical support to help students working from home.''} Students can face technical challenges at home due to a lack of resources and support. Highlighting the importance of technology in a remote research setting, it is necessary to provide more human support and more training than what they received in using technology effectively and fulfilling their project goals. Students need more technical support to practice research and learn physics content. The training documentation including manuals, software handouts, and resources could be available online before the program starts. In addition, to facilitate REU students' engagement with technology and resources immediately upon entry into the REU program, having a point of contact who can respond quickly for technical related help (e.g., software installation) would be helpful. For example, this could be a graduate student or a computer support technician.

\textit{Recommendation 2: ``Provide emotional and environmental support to help students working from home.''} Working remotely brings its own set of norms and challenges that may impact students' productivity. Lack of motivation may be intrinsic to a remote research experience, but it's important to identify these challenges in order to mitigate future problems and increase self-motivation. Mentors and REU coordinators need to ensure that mentees have all the physical and social tools they need (e.g., dedicated office space, a good chair, and good communication technology to facilitate communication) to make progress and handle the isolation challenges. In addition, mentors and REU coordinators can plan social activities outside of their research projects such as an online movie night, a book club, an online yoga class, online game-night, or other events so students can share their daily life together. Mentors and coordinators could offer peer check-in opportunities or schedule times that REU students can work together in the same Zoom room. Students need to support and encourage each other when facing similar challenges. These informal interactions could make students feel connected to their community and increase their motivation.

\textit{Recommendation 3: ``Define clusters of planning skills that are important when doing research remotely.''}
Mentors and REU coordinators can develop a mentoring strategy for helping their students to move forward in their research experience in addition to providing them opportunities to gain enough new content knowledge. Mentors could help mentees to organize their resources. For example, mentors could differentiate between resources that are essential to achieve the project goals and which are optional. In addition, they can schedule more short check-in meetings. Sometimes, having a scheduled meeting at a certain time helps students prepare to work around it and manage their time. Mentors can set out clear goals and deadlines and help mentees divide their research work into manageable chunks. This guidance will help students not get overwhelmed by the extensive amount of new knowledge and what they have to do. %Besides, Program coordinators could carefully well-designed goals of research programs that have the potential to illuminate feelings of frustration.

%Mentors and REU coordinators need to ensure that they provide support for their mentees, such as helping their students to identify the essential knowledge and developing a timeline for their project and assisting them to complete each project step by being approachable and available to them. 

%up to here

\textit{Recommendation 4: ``Provide communication opportunities among REU students.''} Mentors can require each student to make a presentation at a weekly seminar, with one or two faculty assigned and all the other REU students to ask questions and provide feedback. Students need more communication in order to get familiar with the other community members and the broad range of research they are doing. Mentors and REU coordinators can create structured activities to facilitate REU student interactions. 

As a part of this community experience in the remote REU setting, mentors should help mentees to strengthen their peer community and get them in touch with other mentees in the program. For instance, ask all students to gather for lunch once a week to discuss various aspects of remote research and life events. However, this recommendation takes time to build trust among students. Mentors also can ask students to develop and share with peers a work plan that includes short-term and long-term goals for their projects and provide each other feedback. Building these communities may impact students' future career decision. We recommend mentors ask students to work on a common goal through creative activities and communication. For instance, making an REU T-shirt requires lots of group work and informal communication. The graduate school preparation sessions could be served as another form of opportunity that students can share their resources, support each other and exchange their learning experiences. 

\textit{Recommendation 5: ``Find the best technology to communicate with students.''} It is important for both mentors and mentees to contribute to the new norm by keeping communication ways open. Mentors can define a cluster of social skills that are important in doing research remotely. For instance, mentors and mentees can discuss these questions: how often can the student meet their mentors? How often can the student expect to receive feedback for their progress? How long will the meeting be? How should mentees document their progress? Learning about these norms may also benefit students who want to go to graduate school and provide an opportunity for them to learn how to communicate with their Ph.D. advisor. 

\textit{Recommendation 6: ``Discuss about expectations and goals.''}
It is important to understand the mentees' previous educational experiences, future career goals, and their REU goals, and identify a potential barrier to these goals. For instance, the programs can be explicit about their goals, so students understand what they are getting into.%mentors can use a pre-survey to evaluate students' goals of participating in the REU program and then structure opportunities to meet those needs. 
Besides, mentors can be open to hearing students' experiences and perspectives around doing research. Mentors and mentees need to communicate clearly from the start about their roles, expectations, and responsibilities. For instance, mentors can define project goals and expected outcomes at the start of the REU programs. We believe these communication opportunities help students to take more ownership of their projects and their learning.
%We believe if mentors and REU coordinators prioritize one goal over another when confronted with conflicting goals and think of new procedures and policies such as more social hours, peer feedback, and reflection time, and maintain frequent contact with mentees to make sure the REU experience runs smoothly, this issue could be resolved.

%Although COVID-19 and the remote format created some challenges for REU participants and their mentors. Some aspects of the experience were beneficial for mentors. For example, no travel, pre-trip orientation, and housing costs, and less effort are required to organize meetings. Besides, some mentors found benefits in adapting their experimental projects to have a more computational focus, which they may retain even when in-person, hands-on opportunities return in future years of the REU.

%There is still much left to explore and study regarding undergraduate research experiences. There are opportunities to compare students who have access to the remote research labs with those participating in in-person research labs, larger samples with more diverse demographics, and more detailed information on the underlying mechanisms, including how and why these outcomes happened, and how short-term outcomes (e.g., increased interest) relate to long-term outcomes (e.g., retention in STEM majors or entering a STEM Ph.D. program or career). 

In summary, we hope our study provides the PER community with an insightful introduction to the challenges and different outcomes of remote REU programs. We also hope this study provides encouragement and practical advice for REU mentors and coordinators to design programs that foster the best outcomes and avoid likely pitfalls.
%Second, to what extent do remote research programs allow students to engage in research successfully? Future studies need to seek more information on the underlying mechanisms, including how and why these outcomes happened, which mediators (e.g., norms, community, tools) support particular outcomes, and how these short-term outcomes increase long-term retention in STEM majors. Lastly, in examining the benefit of the remote research programs more research is needed to uncover more specific practices and support structures that help students develop the skills necessary to succeed in their future career. 
\section{ACKNOWLEDGEMENT}
The authors would like to thank all participants who shared their REU experiences. Without their consistent participation during the early months of the pandemic this work would not have been possible. This work was supported by the National Science Foundation under Grant No. 1846321.

\appendix
\renewcommand{\thesection}{\Alph{section}.\arabic{section}}
\setcounter{section}{0}
\begin{appendices}
\section{Projects' characteristics}
 %\section*{APPENDIX: REU Projects' characteristics}
\begin{table*}[tbh]
\caption{Projects' characteristics.}
\label{tab: Project}
\begin{tabular}{l|l|l}
\hline\hline
{Name}  & {Area of research} 
& {Description of project} \\
\hline 
{Andrew} & 
{Nuclear Physics} &
{Understand the efficiency of detector, learn about details of the old nuclear}\\
{} &{} & {reaction simulations, and refine the new simulation}\\

{Bruce} &
{Quantum Nonlinear}   
 & {Numerically model quantum optical devices, learn PyBoard coding, use digital}\\
{} &{optics} & {time delay and construct the circuit with equipment that shipped to his home} \\

{Caleb} & 
{Atomic Physics} 
 & {Examine at the atomic structure and different spectroscopies and make a model for}\\
 {} &{} & {specific properties}\\

{David} & 
{Acoustic} 
 & {Make a basic resonator model to learn the modeling program and then make an}\\
 {} &{} & {acoustic model of a reed instrument} \\
 
{Emma} & 
{Physics Education}
 & {She had two projects: 1-PER: science outreach and 2- Build a 3D-printed particle}\\
 {} &{and Particle Physics} & {trap and work with electronics that shipped to her home to make circuits}\\

{Frieda} & 
{High Energy Physics} 
 & {Use different models in high-energy physics to predict the probability of different}\\
 {} &{} & {decay modes in collisions} \\

{Grace} & 
{Solid State Physics} 
 & {Learn the density functional theory and model certain molecules and look at the}\\
 {} &{} & {dynamics of the system}\\

{Helen} & 
{Nuclear Physics} 
 & {Simulate the decay process of short lived isotopes, learn about different isotopes}\\
 {} &{} & {and different spectra fields, and literature half lives}\\

{Ivan} 	&  %yusen
{Nuclear Physics} 
	& {Gain knowledge about CMS and LHC and use simulations and experimental data to}\\
 {} &{} & {refine the codes for detection of the charged particles in a large collider experiment}\\

{Joshua} & 
{Nuclear Physics} 
 &  {Understand neutron mirror model and add new equations into the old code to solve}\\
 {} &{} & {problems related to nuclear physics}\\
 \hline\hline

  \end{tabular}
\end{table*}
\end{appendices}
\nocite{*}
\bibliographystyle{apsrev4-1}
\bibliography{RQ1.bib}

\begin{thebibliography}{74}%
\makeatletter
\providecommand \@ifxundefined [1]{%
 \@ifx{#1\undefined}
}%
\providecommand \@ifnum [1]{%
 \ifnum #1\expandafter \@firstoftwo
 \else \expandafter \@secondoftwo
 \fi
}%
\providecommand \@ifx [1]{%
 \ifx #1\expandafter \@firstoftwo
 \else \expandafter \@secondoftwo
 \fi
}%
\providecommand \natexlab [1]{#1}%
\providecommand \enquote  [1]{``#1''}%
\providecommand \bibnamefont  [1]{#1}%
\providecommand \bibfnamefont [1]{#1}%
\providecommand \citenamefont [1]{#1}%
\providecommand \href@noop [0]{\@secondoftwo}%
\providecommand \href [0]{\begingroup \@sanitize@url \@href}%
\providecommand \@href[1]{\@@startlink{#1}\@@href}%
\providecommand \@@href[1]{\endgroup#1\@@endlink}%
\providecommand \@sanitize@url [0]{\catcode `\\12\catcode `\$12\catcode
  `\&12\catcode `\#12\catcode `\^12\catcode `\_12\catcode `\%12\relax}%
\providecommand \@@startlink[1]{}%
\providecommand \@@endlink[0]{}%
\providecommand \url  [0]{\begingroup\@sanitize@url \@url }%
\providecommand \@url [1]{\endgroup\@href {#1}{\urlprefix }}%
\providecommand \urlprefix  [0]{URL }%
\providecommand \Eprint [0]{\href }%
\providecommand \doibase [0]{http://dx.doi.org/}%
\providecommand \selectlanguage [0]{\@gobble}%
\providecommand \bibinfo  [0]{\@secondoftwo}%
\providecommand \bibfield  [0]{\@secondoftwo}%
\providecommand \translation [1]{[#1]}%
\providecommand \BibitemOpen [0]{}%
\providecommand \bibitemStop [0]{}%
\providecommand \bibitemNoStop [0]{.\EOS\space}%
\providecommand \EOS [0]{\spacefactor3000\relax}%
\providecommand \BibitemShut  [1]{\csname bibitem#1\endcsname}%
\let\auto@bib@innerbib\@empty
%</preamble>
\bibitem [{\citenamefont {Council}(1999)}]{council_transforming_1999}%
  \BibitemOpen
  \bibfield  {author} {\bibinfo {author} {\bibfnamefont {N.~R.}\ \bibnamefont
  {Council}},\ }\href@noop {} {\emph {\bibinfo {title} {Transforming
  Undergraduate Education in Science, Mathematics, Engineering, and
  Technology}}}\ (\bibinfo  {publisher} {The National Academies Press},\
  \bibinfo {address} {Washington, DC},\ \bibinfo {year} {1999})\BibitemShut
  {NoStop}%
\bibitem [{\citenamefont {Council}(2002)}]{council_improving_2002}%
  \BibitemOpen
  \bibfield  {author} {\bibinfo {author} {\bibfnamefont {N.~R.}\ \bibnamefont
  {Council}},\ }\href {\doibase https://doi.org/10.17226/10711} {\emph
  {\bibinfo {title} {Improving undergraduate instruction in Science,
  Technology, Engineering, and Mathematics: report of a workshop}}}\ (\bibinfo
  {publisher} {The National Academies Press},\ \bibinfo {address} {Washington,
  DC},\ \bibinfo {year} {2002})\BibitemShut {NoStop}%
\bibitem [{\citenamefont {Wenzel}\ and\ \citenamefont
  {Karukstis}(2004)}]{wenzel_enhancing_2004}%
  \BibitemOpen
  \bibfield  {author} {\bibinfo {author} {\bibfnamefont {T.~J.}\ \bibnamefont
  {Wenzel}}\ and\ \bibinfo {author} {\bibfnamefont {K.~K.}\ \bibnamefont
  {Karukstis}},\ }\href {\doibase https://doi.org/10.1021/ed081p468} {\bibfield
   {journal} {\bibinfo  {journal} {J. Chem. Educ.}\ }\textbf {\bibinfo {volume}
  {81}},\ \bibinfo {pages} {468} (\bibinfo {year} {2004})},\ \bibinfo {note}
  {publisher: American Chemical Society}\BibitemShut {NoStop}%
\bibitem [{\citenamefont {Kenny}\ \emph {et~al.}(2001)\citenamefont {Kenny},
  \citenamefont {Thomas}, \citenamefont {Katkin}, \citenamefont {Lemming},
  \citenamefont {Smith}, \citenamefont {Glaser},\ and\ \citenamefont
  {Gross}}]{Kenny_2001}%
  \BibitemOpen
  \bibfield  {author} {\bibinfo {author} {\bibfnamefont {S.~S.}\ \bibnamefont
  {Kenny}}, \bibinfo {author} {\bibfnamefont {E.}~\bibnamefont {Thomas}},
  \bibinfo {author} {\bibfnamefont {W.}~\bibnamefont {Katkin}}, \bibinfo
  {author} {\bibfnamefont {M.}~\bibnamefont {Lemming}}, \bibinfo {author}
  {\bibfnamefont {P.}~\bibnamefont {Smith}}, \bibinfo {author} {\bibfnamefont
  {M.}~\bibnamefont {Glaser}}, \ and\ \bibinfo {author} {\bibfnamefont
  {W.}~\bibnamefont {Gross}},\ }\href@noop {} {\emph {\bibinfo {title}
  {Reinventing undergraduate education: three years after the Boyer report}}}\
  (\bibinfo  {publisher} {Stony Brook University: Office of the President},\
  \bibinfo {address} {NY},\ \bibinfo {year} {2001})\BibitemShut {NoStop}%
\bibitem [{\citenamefont {Kuh}(2008)}]{kuh_high-impact_2008}%
  \BibitemOpen
  \bibfield  {author} {\bibinfo {author} {\bibfnamefont {G.~D.}\ \bibnamefont
  {Kuh}},\ }\href@noop {} {\emph {\bibinfo {title} {High-impact educational
  practices: what they are, who has access to them, and why they matter}}}\
  (\bibinfo  {publisher} {Association of American Colleges and Universities},\
  \bibinfo {address} {Washington, D.C.},\ \bibinfo {year} {2008})\BibitemShut
  {NoStop}%
\bibitem [{\citenamefont {Kardash}(2000)}]{kardash_evaluation_2000}%
  \BibitemOpen
  \bibfield  {author} {\bibinfo {author} {\bibfnamefont {C.~M.}\ \bibnamefont
  {Kardash}},\ }\href {\doibase https://doi.org/10.1037/0022-0663.92.1.191}
  {\bibfield  {journal} {\bibinfo  {journal} {J. Educ. Psychol.}\ }\textbf
  {\bibinfo {volume} {92}},\ \bibinfo {pages} {191} (\bibinfo {year}
  {2000})}\BibitemShut {NoStop}%
\bibitem [{\citenamefont {Lopatto}(2007)}]{lopatto_undergraduate_2007}%
  \BibitemOpen
  \bibfield  {author} {\bibinfo {author} {\bibfnamefont {D.}~\bibnamefont
  {Lopatto}},\ }\href {\doibase https://doi.org/10.1187/cbe.07-06-0039}
  {\bibfield  {journal} {\bibinfo  {journal} {CBE Life Sci. Educ.}\ }\textbf
  {\bibinfo {volume} {6}},\ \bibinfo {pages} {297} (\bibinfo {year}
  {2007})}\BibitemShut {NoStop}%
\bibitem [{\citenamefont {Hathaway}\ \emph {et~al.}(2002)\citenamefont
  {Hathaway}, \citenamefont {Nagda},\ and\ \citenamefont
  {Gregerman}}]{hathaway_relationship_2002}%
  \BibitemOpen
  \bibfield  {author} {\bibinfo {author} {\bibfnamefont {R.~S.}\ \bibnamefont
  {Hathaway}}, \bibinfo {author} {\bibfnamefont {B.~R.~A.}\ \bibnamefont
  {Nagda}}, \ and\ \bibinfo {author} {\bibfnamefont {S.~R.}\ \bibnamefont
  {Gregerman}},\ }\href@noop {} {\bibfield  {journal} {\bibinfo  {journal} {J.
  Coll. Stud. Dev.}\ }\textbf {\bibinfo {volume} {43}},\ \bibinfo {pages} {614}
  (\bibinfo {year} {2002})}\BibitemShut {NoStop}%
\bibitem [{\citenamefont {Seymour}\ \emph {et~al.}(2004)\citenamefont
  {Seymour}, \citenamefont {Hunter}, \citenamefont {Laursen},\ and\
  \citenamefont {DeAntoni}}]{seymour_establishing_2004}%
  \BibitemOpen
  \bibfield  {author} {\bibinfo {author} {\bibfnamefont {E.}~\bibnamefont
  {Seymour}}, \bibinfo {author} {\bibfnamefont {A.-B.}\ \bibnamefont {Hunter}},
  \bibinfo {author} {\bibfnamefont {S.~L.}\ \bibnamefont {Laursen}}, \ and\
  \bibinfo {author} {\bibfnamefont {T.}~\bibnamefont {DeAntoni}},\ }\href@noop
  {} {\bibfield  {journal} {\bibinfo  {journal} {Sci. Educ.}\ }\textbf
  {\bibinfo {volume} {88}} (\bibinfo {year} {2004})}\BibitemShut {NoStop}%
\bibitem [{\citenamefont {Lopatto}(2004)}]{lopatto_survey_2004}%
  \BibitemOpen
  \bibfield  {author} {\bibinfo {author} {\bibfnamefont {D.}~\bibnamefont
  {Lopatto}},\ }\href {\doibase https://doi.org/10.1187/cbe.04-07-0045}
  {\bibfield  {journal} {\bibinfo  {journal} {CBE Life Sci. Educ.}\ }\textbf
  {\bibinfo {volume} {3}},\ \bibinfo {pages} {270} (\bibinfo {year}
  {2004})}\BibitemShut {NoStop}%
\bibitem [{\citenamefont {Hunter}\ \emph {et~al.}(2007)\citenamefont {Hunter},
  \citenamefont {Laursen},\ and\ \citenamefont
  {Seymour}}]{hunter_becoming_2007}%
  \BibitemOpen
  \bibfield  {author} {\bibinfo {author} {\bibfnamefont {A.-B.}\ \bibnamefont
  {Hunter}}, \bibinfo {author} {\bibfnamefont {S.~L.}\ \bibnamefont {Laursen}},
  \ and\ \bibinfo {author} {\bibfnamefont {E.}~\bibnamefont {Seymour}},\
  }\href@noop {} {\bibfield  {journal} {\bibinfo  {journal} {Sci. Educ.}\
  }\textbf {\bibinfo {volume} {91}},\ \bibinfo {pages} {36} (\bibinfo {year}
  {2007})}\BibitemShut {NoStop}%
\bibitem [{\citenamefont {Laursen}\ \emph {et~al.}(2010)\citenamefont
  {Laursen}, \citenamefont {Hunter}, \citenamefont {Seymour}, \citenamefont
  {Thiry},\ and\ \citenamefont {Melton}}]{laursen_undergraduate_2010}%
  \BibitemOpen
  \bibfield  {author} {\bibinfo {author} {\bibfnamefont {S.}~\bibnamefont
  {Laursen}}, \bibinfo {author} {\bibfnamefont {A.-B.}\ \bibnamefont {Hunter}},
  \bibinfo {author} {\bibfnamefont {E.}~\bibnamefont {Seymour}}, \bibinfo
  {author} {\bibfnamefont {H.}~\bibnamefont {Thiry}}, \ and\ \bibinfo {author}
  {\bibfnamefont {G.}~\bibnamefont {Melton}},\ }\href@noop {} {\emph {\bibinfo
  {title} {Undergraduate research in the sciences: engaging students in real
  science}}}\ (\bibinfo  {publisher} {Jossey-Bass},\ \bibinfo {year}
  {2010})\BibitemShut {NoStop}%
\bibitem [{\citenamefont {Estrada}\ \emph {et~al.}(2011)\citenamefont
  {Estrada}, \citenamefont {Woodcock}, \citenamefont {Hernandez},\ and\
  \citenamefont {Schultz}}]{estrada_toward_2011}%
  \BibitemOpen
  \bibfield  {author} {\bibinfo {author} {\bibfnamefont {M.}~\bibnamefont
  {Estrada}}, \bibinfo {author} {\bibfnamefont {A.}~\bibnamefont {Woodcock}},
  \bibinfo {author} {\bibfnamefont {P.~R.}\ \bibnamefont {Hernandez}}, \ and\
  \bibinfo {author} {\bibfnamefont {P.~W.}\ \bibnamefont {Schultz}},\ }\href
  {\doibase https://doi.org/10.1037/a0020743} {\bibfield  {journal} {\bibinfo
  {journal} {J. Educ. Psychol.}\ }\textbf {\bibinfo {volume} {103}},\ \bibinfo
  {pages} {206} (\bibinfo {year} {2011})}\BibitemShut {NoStop}%
\bibitem [{\citenamefont {Graham}\ \emph {et~al.}(2013)\citenamefont {Graham},
  \citenamefont {Frederick}, \citenamefont {Byars-Winston}, \citenamefont
  {Hunter},\ and\ \citenamefont {Handelsman}}]{graham_increasing_2013}%
  \BibitemOpen
  \bibfield  {author} {\bibinfo {author} {\bibfnamefont {M.~J.}\ \bibnamefont
  {Graham}}, \bibinfo {author} {\bibfnamefont {J.}~\bibnamefont {Frederick}},
  \bibinfo {author} {\bibfnamefont {A.}~\bibnamefont {Byars-Winston}}, \bibinfo
  {author} {\bibfnamefont {A.-B.}\ \bibnamefont {Hunter}}, \ and\ \bibinfo
  {author} {\bibfnamefont {J.}~\bibnamefont {Handelsman}},\ }\href {\doibase
  https://doi.org/10.1126/science.1240487} {\bibfield  {journal} {\bibinfo
  {journal} {Science}\ }\textbf {\bibinfo {volume} {341}},\ \bibinfo {pages}
  {1455} (\bibinfo {year} {2013})}\BibitemShut {NoStop}%
\bibitem [{\citenamefont {Hausmann}\ \emph {et~al.}(2007)\citenamefont
  {Hausmann}, \citenamefont {Schofield},\ and\ \citenamefont
  {Woods}}]{hausmann_sense_2007}%
  \BibitemOpen
  \bibfield  {author} {\bibinfo {author} {\bibfnamefont {L.~R.~M.}\
  \bibnamefont {Hausmann}}, \bibinfo {author} {\bibfnamefont {J.~W.}\
  \bibnamefont {Schofield}}, \ and\ \bibinfo {author} {\bibfnamefont {R.~L.}\
  \bibnamefont {Woods}},\ }\href@noop {} {\bibfield  {journal} {\bibinfo
  {journal} {Res. High. Educ.}\ }\textbf {\bibinfo {volume} {48}},\ \bibinfo
  {pages} {803} (\bibinfo {year} {2007})}\BibitemShut {NoStop}%
\bibitem [{\citenamefont {Dolan}\ and\ \citenamefont
  {Johnson}(2009)}]{dolan_toward_2009}%
  \BibitemOpen
  \bibfield  {author} {\bibinfo {author} {\bibfnamefont {E.}~\bibnamefont
  {Dolan}}\ and\ \bibinfo {author} {\bibfnamefont {D.}~\bibnamefont
  {Johnson}},\ }\href {\doibase https://doi.org/10.1007/s10956-009-9165-3}
  {\bibfield  {journal} {\bibinfo  {journal} {J. Sci. Educ. Technol.}\ }\textbf
  {\bibinfo {volume} {18}},\ \bibinfo {pages} {487} (\bibinfo {year}
  {2009})}\BibitemShut {NoStop}%
\bibitem [{\citenamefont {Eagan}\ \emph {et~al.}(2013)\citenamefont {Eagan},
  \citenamefont {Hurtado}, \citenamefont {Chang}, \citenamefont {Garcia},
  \citenamefont {Herrera},\ and\ \citenamefont {Garibay}}]{eagan_making_2013}%
  \BibitemOpen
  \bibfield  {author} {\bibinfo {author} {\bibfnamefont {M.~K.}\ \bibnamefont
  {Eagan}}, \bibinfo {author} {\bibfnamefont {S.}~\bibnamefont {Hurtado}},
  \bibinfo {author} {\bibfnamefont {M.~J.}\ \bibnamefont {Chang}}, \bibinfo
  {author} {\bibfnamefont {G.~A.}\ \bibnamefont {Garcia}}, \bibinfo {author}
  {\bibfnamefont {F.~A.}\ \bibnamefont {Herrera}}, \ and\ \bibinfo {author}
  {\bibfnamefont {J.~C.}\ \bibnamefont {Garibay}},\ }\href {\doibase
  https://doi.org/10.3102/0002831213482038} {\bibfield  {journal} {\bibinfo
  {journal} {Am. Educ. Res. J.}\ }\textbf {\bibinfo {volume} {50}},\ \bibinfo
  {pages} {683} (\bibinfo {year} {2013})}\BibitemShut {NoStop}%
\bibitem [{\citenamefont {of~Sciences}\ \emph {et~al.}(2017)\citenamefont
  {of~Sciences}, \citenamefont {Medicine},\ and\ \citenamefont
  {Engineering}}]{national_academies_of_sciences_undergraduate_2017}%
  \BibitemOpen
  \bibfield  {author} {\bibinfo {author} {\bibfnamefont {N.~A.}\ \bibnamefont
  {of~Sciences}}, \bibinfo {author} {\bibnamefont {Medicine}}, \ and\ \bibinfo
  {author} {\bibnamefont {Engineering}},\ }\href@noop {} {\emph {\bibinfo
  {title} {Undergraduate Research Experiences for STEM students: successes,
  challenges, and opportunities}}},\ edited by\ \bibinfo {editor}
  {\bibfnamefont {J.}~\bibnamefont {Gentile}}, \bibinfo {editor} {\bibfnamefont
  {K.}~\bibnamefont {Brenner}}, \ and\ \bibinfo {editor} {\bibfnamefont
  {A.}~\bibnamefont {Stephens}}\ (\bibinfo  {publisher} {The National Academies
  Press},\ \bibinfo {address} {Washington, DC},\ \bibinfo {year}
  {2017})\BibitemShut {NoStop}%
\bibitem [{\citenamefont {Blockus}(2016)}]{blockus_strengthening_2016}%
  \BibitemOpen
  \bibfield  {author} {\bibinfo {author} {\bibfnamefont {L.}~\bibnamefont
  {Blockus}},\ }\href@noop {} {\bibfield  {journal} {\bibinfo  {journal}
  {National Academies of Sciences, Engineering, and Medicine (NASEM)}\ ,\
  \bibinfo {pages} {31}} (\bibinfo {year} {2016})}\BibitemShut {NoStop}%
\bibitem [{\citenamefont {Russell}\ \emph
  {et~al.}(2007{\natexlab{a}})\citenamefont {Russell}, \citenamefont
  {Hancock},\ and\ \citenamefont {McCullough}}]{russell_benefits_2007}%
  \BibitemOpen
  \bibfield  {author} {\bibinfo {author} {\bibfnamefont {S.~H.}\ \bibnamefont
  {Russell}}, \bibinfo {author} {\bibfnamefont {M.~P.}\ \bibnamefont
  {Hancock}}, \ and\ \bibinfo {author} {\bibfnamefont {J.}~\bibnamefont
  {McCullough}},\ }\href@noop {} {\bibfield  {journal} {\bibinfo  {journal}
  {Science}\ }\textbf {\bibinfo {volume} {316}},\ \bibinfo {pages} {548}
  (\bibinfo {year} {2007}{\natexlab{a}})}\BibitemShut {NoStop}%
\bibitem [{\citenamefont {Forrester}(2021)}]{forrester_how_2021}%
  \BibitemOpen
  \bibfield  {author} {\bibinfo {author} {\bibfnamefont {N.}~\bibnamefont
  {Forrester}},\ }\href {\doibase DOI: 10.1038/d41586-021-01209-2} {\bibfield
  {journal} {\bibinfo  {journal} {Nature}\ } (\bibinfo {year} {2021}),\ DOI:
  10.1038/d41586-021-01209-2}\BibitemShut {NoStop}%
\bibitem [{\citenamefont {Leont'ev}(1974)}]{leontev_problem_1974}%
  \BibitemOpen
  \bibfield  {author} {\bibinfo {author} {\bibfnamefont {A.~N.}\ \bibnamefont
  {Leont'ev}},\ }\href@noop {} {\bibfield  {journal} {\bibinfo  {journal}
  {Soviet Psychology}\ }\textbf {\bibinfo {volume} {13}},\ \bibinfo {pages} {4}
  (\bibinfo {year} {1974})},\ \bibinfo {note} {publisher:
  Routledge}\BibitemShut {NoStop}%
\bibitem [{\citenamefont {Roth}\ and\ \citenamefont
  {Lee}(2007)}]{roth_vygotskys_2007}%
  \BibitemOpen
  \bibfield  {author} {\bibinfo {author} {\bibfnamefont {W.-M.}\ \bibnamefont
  {Roth}}\ and\ \bibinfo {author} {\bibfnamefont {Y.-J.}\ \bibnamefont {Lee}},\
  }\href@noop {} {\bibfield  {journal} {\bibinfo  {journal} {Rev. Educ. Res.}\
  }\textbf {\bibinfo {volume} {77}},\ \bibinfo {pages} {186} (\bibinfo {year}
  {2007})},\ \bibinfo {note} {publisher: American Educational Research
  Association}\BibitemShut {NoStop}%
\bibitem [{\citenamefont {Engestr{\"o}m}(1987)}]{engestrom_learning_1987}%
  \BibitemOpen
  \bibfield  {author} {\bibinfo {author} {\bibfnamefont {Y.}~\bibnamefont
  {Engestr{\"o}m}},\ }\href@noop {} {\emph {\bibinfo {title} {Learning by
  expanding: A activity-theoretical approach to developmental research}}}\
  (\bibinfo  {publisher} {Cambridge University Press},\ \bibinfo {address} {New
  York, NY},\ \bibinfo {year} {1987})\BibitemShut {NoStop}%
\bibitem [{\citenamefont {Zohrabi-Alaee}\ \emph {et~al.}(2022)\citenamefont
  {Zohrabi-Alaee}, \citenamefont {Campbell},\ and\ \citenamefont
  {Zwickl}}]{zohrabi_alaee_impact_2022}%
  \BibitemOpen
  \bibfield  {author} {\bibinfo {author} {\bibfnamefont {D.}~\bibnamefont
  {Zohrabi-Alaee}}, \bibinfo {author} {\bibfnamefont {M.~K.}\ \bibnamefont
  {Campbell}}, \ and\ \bibinfo {author} {\bibfnamefont {B.~M.}\ \bibnamefont
  {Zwickl}},\ }\href@noop {} {\bibfield  {journal} {\bibinfo  {journal} {Phys.
  Rev. Phys. Educ. Res.}\ }\textbf {\bibinfo {volume} {18}},\ \bibinfo {pages}
  {010101} (\bibinfo {year} {2022})}\BibitemShut {NoStop}%
\bibitem [{\citenamefont {Marton}\ and\ \citenamefont
  {Booth}(1997{\natexlab{a}})}]{Marton1997}%
  \BibitemOpen
  \bibfield  {author} {\bibinfo {author} {\bibfnamefont {F.}~\bibnamefont
  {Marton}}\ and\ \bibinfo {author} {\bibfnamefont {S.~A.}\ \bibnamefont
  {Booth}},\ }\href@noop {} {\emph {\bibinfo {title} {Learning and
  awareness}}}\ (\bibinfo  {publisher} {Mahwah, N.J. : L. Erlbaum Associates},\
  \bibinfo {year} {1997})\BibitemShut {NoStop}%
\bibitem [{\citenamefont {Marton}(2000)}]{marton2000structure}%
  \BibitemOpen
  \bibfield  {author} {\bibinfo {author} {\bibfnamefont {F.}~\bibnamefont
  {Marton}},\ }\href@noop {} {\bibfield  {journal} {\bibinfo  {journal}
  {Phenomenography}\ ,\ \bibinfo {pages} {102}} (\bibinfo {year}
  {2000})}\BibitemShut {NoStop}%
\bibitem [{\citenamefont {Zohrabi-Alaee}\ and\ \citenamefont
  {Zwickl}(2021)}]{Zohrabi_perc2021}%
  \BibitemOpen
  \bibfield  {author} {\bibinfo {author} {\bibfnamefont {D.}~\bibnamefont
  {Zohrabi-Alaee}}\ and\ \bibinfo {author} {\bibfnamefont {B.~M.}\ \bibnamefont
  {Zwickl}},\ }in\ \href@noop {} {\emph {\bibinfo {booktitle} {Physics
  Education Research Conference 2021}}},\ \bibinfo {series and number} {PER
  Conference}\ (\bibinfo {address} {Virtual Conference},\ \bibinfo {year}
  {2021})\ pp.\ \bibinfo {pages} {480--485}\BibitemShut {NoStop}%
\bibitem [{\citenamefont {Helba}\ \emph {et~al.}(2019)\citenamefont {Helba},
  \citenamefont {Porter}, \citenamefont {Nicholson},\ and\ \citenamefont
  {Ivie}}]{AIP2019}%
  \BibitemOpen
  \bibfield  {author} {\bibinfo {author} {\bibfnamefont {C.}~\bibnamefont
  {Helba}}, \bibinfo {author} {\bibfnamefont {A.~M.}\ \bibnamefont {Porter}},
  \bibinfo {author} {\bibfnamefont {S.}~\bibnamefont {Nicholson}}, \ and\
  \bibinfo {author} {\bibfnamefont {R.}~\bibnamefont {Ivie}},\ }\href
  {https://www.aip.org/statistics/reports/women-among-physics-and-astronomy-faculty}
  {\enquote {\bibinfo {title} {Women among physics and astronomy faculty
  results from the 2018 academic workforce survey},}\ } (\bibinfo {year}
  {2019})\BibitemShut {NoStop}%
\bibitem [{\citenamefont {Walsh}\ and\ \citenamefont {Tyler}(2022)}]{AIP2022}%
  \BibitemOpen
  \bibfield  {author} {\bibinfo {author} {\bibfnamefont {C.}~\bibnamefont
  {Walsh}}\ and\ \bibinfo {author} {\bibfnamefont {J.}~\bibnamefont {Tyler}},\
  }\href {https://www.aip.org/statistics/reports/covid-faculty-qualityofwork}
  {\enquote {\bibinfo {title} {Self-reported changes in quality of work as a
  result of the covid-19 pandemic for faculty members in physics and
  astronomy},}\ } (\bibinfo {year} {2022})\BibitemShut {NoStop}%
\bibitem [{\citenamefont {Myers}\ \emph {et~al.}(2020)\citenamefont {Myers},
  \citenamefont {Tham}, \citenamefont {Yin}, \citenamefont {Cohodes},
  \citenamefont {Thursby}, \citenamefont {Thursby}, \citenamefont {Schiffer},
  \citenamefont {Walsh}, \citenamefont {Lakhani},\ and\ \citenamefont
  {Wang}}]{myers2020unequal}%
  \BibitemOpen
  \bibfield  {author} {\bibinfo {author} {\bibfnamefont {K.~R.}\ \bibnamefont
  {Myers}}, \bibinfo {author} {\bibfnamefont {W.~Y.}\ \bibnamefont {Tham}},
  \bibinfo {author} {\bibfnamefont {Y.}~\bibnamefont {Yin}}, \bibinfo {author}
  {\bibfnamefont {N.}~\bibnamefont {Cohodes}}, \bibinfo {author} {\bibfnamefont
  {J.~G.}\ \bibnamefont {Thursby}}, \bibinfo {author} {\bibfnamefont {M.~C.}\
  \bibnamefont {Thursby}}, \bibinfo {author} {\bibfnamefont {P.}~\bibnamefont
  {Schiffer}}, \bibinfo {author} {\bibfnamefont {J.~T.}\ \bibnamefont {Walsh}},
  \bibinfo {author} {\bibfnamefont {K.~R.}\ \bibnamefont {Lakhani}}, \ and\
  \bibinfo {author} {\bibfnamefont {D.}~\bibnamefont {Wang}},\ }\href@noop {}
  {\bibfield  {journal} {\bibinfo  {journal} {Nat. Hum. Behav.}\ }\textbf
  {\bibinfo {volume} {4}},\ \bibinfo {pages} {880} (\bibinfo {year}
  {2020})}\BibitemShut {NoStop}%
\bibitem [{\citenamefont {Krukowski}\ \emph {et~al.}(2021)\citenamefont
  {Krukowski}, \citenamefont {Jagsi},\ and\ \citenamefont
  {Cardel}}]{katz2021re}%
  \BibitemOpen
  \bibfield  {author} {\bibinfo {author} {\bibfnamefont {R.}~\bibnamefont
  {Krukowski}}, \bibinfo {author} {\bibfnamefont {R.}~\bibnamefont {Jagsi}}, \
  and\ \bibinfo {author} {\bibfnamefont {M.}~\bibnamefont {Cardel}},\
  }\href@noop {} {\bibfield  {journal} {\bibinfo  {journal} {J. Womens Health}\
  }\textbf {\bibinfo {volume} {30}},\ \bibinfo {pages} {341} (\bibinfo {year}
  {2021})}\BibitemShut {NoStop}%
\bibitem [{\citenamefont {Deddose}(2018)}]{noauthor_dedoose_2018}%
  \BibitemOpen
  \bibfield  {author} {\bibinfo {author} {\bibnamefont {Deddose}},\ }\href
  {https://app.dedoose.com/App/?Version=8.0.35} {\enquote {\bibinfo {title}
  {Dedoose version 8.0.35, web application for managing, analyzing, and
  presenting qualitative and mixed method research data},}\ } (\bibinfo {year}
  {2018})\BibitemShut {NoStop}%
\bibitem [{\citenamefont {Nagda}\ \emph {et~al.}(1998)\citenamefont {Nagda},
  \citenamefont {Gregerman}, \citenamefont {Jonides}, \citenamefont {von
  Hippel},\ and\ \citenamefont {Lerner}}]{nagda_undergraduate_1998}%
  \BibitemOpen
  \bibfield  {author} {\bibinfo {author} {\bibfnamefont {B.~A.}\ \bibnamefont
  {Nagda}}, \bibinfo {author} {\bibfnamefont {S.~R.}\ \bibnamefont
  {Gregerman}}, \bibinfo {author} {\bibfnamefont {J.}~\bibnamefont {Jonides}},
  \bibinfo {author} {\bibfnamefont {W.}~\bibnamefont {von Hippel}}, \ and\
  \bibinfo {author} {\bibfnamefont {J.~S.}\ \bibnamefont {Lerner}},\ }\href
  {\doibase https://doi.org/10.1353/rhe.1998.0016} {\bibfield  {journal}
  {\bibinfo  {journal} {Rev. High. Ed.}\ }\textbf {\bibinfo {volume} {22}},\
  \bibinfo {pages} {55} (\bibinfo {year} {1998})}\BibitemShut {NoStop}%
\bibitem [{\citenamefont {Rodenbusch}\ \emph {et~al.}(2016)\citenamefont
  {Rodenbusch}, \citenamefont {Hernandez}, \citenamefont {Simmons},\ and\
  \citenamefont {Dolan}}]{rodenbusch_early_2016}%
  \BibitemOpen
  \bibfield  {author} {\bibinfo {author} {\bibfnamefont {S.~E.}\ \bibnamefont
  {Rodenbusch}}, \bibinfo {author} {\bibfnamefont {P.~R.}\ \bibnamefont
  {Hernandez}}, \bibinfo {author} {\bibfnamefont {S.~L.}\ \bibnamefont
  {Simmons}}, \ and\ \bibinfo {author} {\bibfnamefont {E.~L.}\ \bibnamefont
  {Dolan}},\ }\href@noop {} {\bibfield  {journal} {\bibinfo  {journal} {CBE
  Life Sci. Educ.}\ }\textbf {\bibinfo {volume} {15}} (\bibinfo {year}
  {2016})}\BibitemShut {NoStop}%
\bibitem [{\citenamefont {Junge}\ \emph {et~al.}(2010)\citenamefont {Junge},
  \citenamefont {Quiñones}, \citenamefont {Kakietek}, \citenamefont
  {Teodorescu},\ and\ \citenamefont {Marsteller}}]{junge_promoting_2010}%
  \BibitemOpen
  \bibfield  {author} {\bibinfo {author} {\bibfnamefont {B.}~\bibnamefont
  {Junge}}, \bibinfo {author} {\bibfnamefont {C.}~\bibnamefont {Quiñones}},
  \bibinfo {author} {\bibfnamefont {J.}~\bibnamefont {Kakietek}}, \bibinfo
  {author} {\bibfnamefont {D.}~\bibnamefont {Teodorescu}}, \ and\ \bibinfo
  {author} {\bibfnamefont {P.}~\bibnamefont {Marsteller}},\ }\href {\doibase
  https://doi.org/10.1187/cbe.09-08-0057} {\bibfield  {journal} {\bibinfo
  {journal} {CBE Life Sci. Educ.}\ }\textbf {\bibinfo {volume} {9}},\ \bibinfo
  {pages} {119} (\bibinfo {year} {2010})}\BibitemShut {NoStop}%
\bibitem [{\citenamefont {Vygotsky}\ \emph {et~al.}(1978)\citenamefont
  {Vygotsky}, \citenamefont {Cole}, \citenamefont {John-Steiner}, \citenamefont
  {Scribner},\ and\ \citenamefont {Souberman}}]{vygotsky_mind_1978}%
  \BibitemOpen
  \bibfield  {author} {\bibinfo {author} {\bibfnamefont {L.~S.}\ \bibnamefont
  {Vygotsky}}, \bibinfo {author} {\bibfnamefont {M.}~\bibnamefont {Cole}},
  \bibinfo {author} {\bibfnamefont {V.}~\bibnamefont {John-Steiner}}, \bibinfo
  {author} {\bibfnamefont {S.}~\bibnamefont {Scribner}}, \ and\ \bibinfo
  {author} {\bibfnamefont {E.}~\bibnamefont {Souberman}},\ }\href@noop {}
  {\emph {\bibinfo {title} {Mind in society: development of higher
  psychological processes}}}\ (\bibinfo  {publisher} {Harvard University
  Press},\ \bibinfo {year} {1978})\BibitemShut {NoStop}%
\bibitem [{\citenamefont {Gorgi}(2009)}]{Gorgi_2009}%
  \BibitemOpen
  \bibfield  {author} {\bibinfo {author} {\bibfnamefont {A.}~\bibnamefont
  {Gorgi}},\ }\href@noop {} {\emph {\bibinfo {title} {The descriptive
  phenomenological method in psychology: a modified Husserlian approach}}}\
  (\bibinfo  {publisher} {Duquesne University},\ \bibinfo {address} {Pittsburg,
  PA},\ \bibinfo {year} {2009})\BibitemShut {NoStop}%
\bibitem [{\citenamefont {Groenewald}(2004)}]{Groenewald_2004}%
  \BibitemOpen
  \bibfield  {author} {\bibinfo {author} {\bibfnamefont {T.}~\bibnamefont
  {Groenewald}},\ }\href@noop {} {\bibfield  {journal} {\bibinfo  {journal}
  {Int. J. Qual. Methods}\ }\textbf {\bibinfo {volume} {3}},\ \bibinfo {pages}
  {42} (\bibinfo {year} {2004})}\BibitemShut {NoStop}%
\bibitem [{\citenamefont {Engestr{\"o}m}(2001)}]{Engestrom2001}%
  \BibitemOpen
  \bibfield  {author} {\bibinfo {author} {\bibfnamefont {Y.}~\bibnamefont
  {Engestr{\"o}m}},\ }\href@noop {} {\bibfield  {journal} {\bibinfo  {journal}
  {J. Educ. Work.}\ }\textbf {\bibinfo {volume} {14}},\ \bibinfo {pages} {133}
  (\bibinfo {year} {2001})}\BibitemShut {NoStop}%
\bibitem [{\citenamefont {M.~Hubenthal}\ and\ \citenamefont
  {Frassetto}(2007)}]{2007AGUFMED33B1226H}%
  \BibitemOpen
  \bibfield  {author} {\bibinfo {author} {\bibfnamefont {R.~A.}\ \bibnamefont
  {M.~Hubenthal}, \bibfnamefont {J.~Taber}}\ and\ \bibinfo {author}
  {\bibfnamefont {A.}~\bibnamefont {Frassetto}},\ }in\ \href@noop {} {\emph
  {\bibinfo {booktitle} {Fall Meeting Abstracts}}},\ Vol.\ \bibinfo {volume}
  {2007}\ (\bibinfo {year} {2007})\ pp.\ \bibinfo {pages}
  {33--1226}\BibitemShut {NoStop}%
\bibitem [{\citenamefont {Pascarella}\ and\ \citenamefont
  {Terenzini}(2005)}]{pascarella_how_2005}%
  \BibitemOpen
  \bibfield  {author} {\bibinfo {author} {\bibfnamefont {E.~T.}\ \bibnamefont
  {Pascarella}}\ and\ \bibinfo {author} {\bibfnamefont {P.~T.}\ \bibnamefont
  {Terenzini}},\ }\href@noop {} {\emph {\bibinfo {title} {How College Affects
  Students: a third decade of research}}},\ Vol.~\bibinfo {volume} {2}\
  (\bibinfo  {publisher} {Jossey-Bass},\ \bibinfo {address} {San Francisco,
  CA},\ \bibinfo {year} {2005})\BibitemShut {NoStop}%
\bibitem [{\citenamefont {Pascarella}\ and\ \citenamefont
  {Terenzini}(1991)}]{pascarella_how_1991}%
  \BibitemOpen
  \bibfield  {author} {\bibinfo {author} {\bibfnamefont {E.~T.}\ \bibnamefont
  {Pascarella}}\ and\ \bibinfo {author} {\bibfnamefont {P.~T.}\ \bibnamefont
  {Terenzini}},\ }\href@noop {} {\emph {\bibinfo {title} {How college affects
  students: findings and insights from twenty years of research}}}\ (\bibinfo
  {publisher} {Jossey-Bass Inc.},\ \bibinfo {address} {San Francisco, CA},\
  \bibinfo {year} {1991})\BibitemShut {NoStop}%
\bibitem [{\citenamefont {Lopatto}\ \emph {et~al.}(2008)\citenamefont
  {Lopatto}, \citenamefont {Alvarez}, \citenamefont {Barnard}, \citenamefont
  {Chandrasekaran}, \citenamefont {Chung}, \citenamefont {Du}, \citenamefont
  {Eckdahl}, \citenamefont {Goodman}, \citenamefont {Hauser}, \citenamefont
  {Jones}, \citenamefont {Kopp}, \citenamefont {Kuleck}, \citenamefont
  {McNeil}, \citenamefont {Morris}, \citenamefont {Myka}, \citenamefont
  {Nagengast}, \citenamefont {Overvoorde}, \citenamefont {Poet}, \citenamefont
  {Reed}, \citenamefont {Regisford}, \citenamefont {Revie}, \citenamefont
  {Rosenwald}, \citenamefont {Saville}, \citenamefont {Shaw}, \citenamefont
  {Skuse}, \citenamefont {Smith}, \citenamefont {Smith}, \citenamefont
  {Spratt}, \citenamefont {Stamm}, \citenamefont {Thompson}, \citenamefont
  {Wilson}, \citenamefont {Witkowski}, \citenamefont {Youngblom}, \citenamefont
  {Leung}, \citenamefont {Shaffer}, \citenamefont {Buhler}, \citenamefont
  {Mardis},\ and\ \citenamefont {Elgin}}]{lopatto_undergraduate_2008}%
  \BibitemOpen
  \bibfield  {author} {\bibinfo {author} {\bibfnamefont {D.}~\bibnamefont
  {Lopatto}}, \bibinfo {author} {\bibfnamefont {C.}~\bibnamefont {Alvarez}},
  \bibinfo {author} {\bibfnamefont {D.}~\bibnamefont {Barnard}}, \bibinfo
  {author} {\bibfnamefont {C.}~\bibnamefont {Chandrasekaran}}, \bibinfo
  {author} {\bibfnamefont {H.~M.}\ \bibnamefont {Chung}}, \bibinfo {author}
  {\bibfnamefont {C.}~\bibnamefont {Du}}, \bibinfo {author} {\bibfnamefont
  {T.}~\bibnamefont {Eckdahl}}, \bibinfo {author} {\bibfnamefont {A.~L.}\
  \bibnamefont {Goodman}}, \bibinfo {author} {\bibfnamefont {C.}~\bibnamefont
  {Hauser}}, \bibinfo {author} {\bibfnamefont {C.~J.}\ \bibnamefont {Jones}},
  \bibinfo {author} {\bibfnamefont {O.~R.}\ \bibnamefont {Kopp}}, \bibinfo
  {author} {\bibfnamefont {G.~A.}\ \bibnamefont {Kuleck}}, \bibinfo {author}
  {\bibfnamefont {G.}~\bibnamefont {McNeil}}, \bibinfo {author} {\bibfnamefont
  {R.}~\bibnamefont {Morris}}, \bibinfo {author} {\bibfnamefont {J.~L.}\
  \bibnamefont {Myka}}, \bibinfo {author} {\bibfnamefont {A.}~\bibnamefont
  {Nagengast}}, \bibinfo {author} {\bibfnamefont {P.~J.}\ \bibnamefont
  {Overvoorde}}, \bibinfo {author} {\bibfnamefont {J.~L.}\ \bibnamefont
  {Poet}}, \bibinfo {author} {\bibfnamefont {K.}~\bibnamefont {Reed}}, \bibinfo
  {author} {\bibfnamefont {G.}~\bibnamefont {Regisford}}, \bibinfo {author}
  {\bibfnamefont {D.}~\bibnamefont {Revie}}, \bibinfo {author} {\bibfnamefont
  {A.}~\bibnamefont {Rosenwald}}, \bibinfo {author} {\bibfnamefont
  {K.}~\bibnamefont {Saville}}, \bibinfo {author} {\bibfnamefont
  {M.}~\bibnamefont {Shaw}}, \bibinfo {author} {\bibfnamefont {G.~R.}\
  \bibnamefont {Skuse}}, \bibinfo {author} {\bibfnamefont {C.}~\bibnamefont
  {Smith}}, \bibinfo {author} {\bibfnamefont {M.}~\bibnamefont {Smith}},
  \bibinfo {author} {\bibfnamefont {M.}~\bibnamefont {Spratt}}, \bibinfo
  {author} {\bibfnamefont {J.}~\bibnamefont {Stamm}}, \bibinfo {author}
  {\bibfnamefont {J.~S.}\ \bibnamefont {Thompson}}, \bibinfo {author}
  {\bibfnamefont {B.~A.}\ \bibnamefont {Wilson}}, \bibinfo {author}
  {\bibfnamefont {C.}~\bibnamefont {Witkowski}}, \bibinfo {author}
  {\bibfnamefont {J.}~\bibnamefont {Youngblom}}, \bibinfo {author}
  {\bibfnamefont {W.}~\bibnamefont {Leung}}, \bibinfo {author} {\bibfnamefont
  {C.~D.}\ \bibnamefont {Shaffer}}, \bibinfo {author} {\bibfnamefont
  {J.}~\bibnamefont {Buhler}}, \bibinfo {author} {\bibfnamefont
  {E.}~\bibnamefont {Mardis}}, \ and\ \bibinfo {author} {\bibfnamefont
  {S.~C.~R.}\ \bibnamefont {Elgin}},\ }\href {\doibase
  https://doi.org/10.1126/science.1165351} {\bibfield  {journal} {\bibinfo
  {journal} {Science}\ }\textbf {\bibinfo {volume} {322}},\ \bibinfo {pages}
  {684} (\bibinfo {year} {2008})}\BibitemShut {NoStop}%
\bibitem [{\citenamefont {Geisinger}\ and\ \citenamefont
  {Raman}(2013)}]{geisinger_why_2013}%
  \BibitemOpen
  \bibfield  {author} {\bibinfo {author} {\bibfnamefont {B.~N.}\ \bibnamefont
  {Geisinger}}\ and\ \bibinfo {author} {\bibfnamefont {D.~R.}\ \bibnamefont
  {Raman}},\ }\href@noop {} {\bibfield  {journal} {\bibinfo  {journal} {Int. J.
  Eng. Educ.}\ }\textbf {\bibinfo {volume} {29}},\ \bibinfo {pages} {914}
  (\bibinfo {year} {2013})}\BibitemShut {NoStop}%
\bibitem [{\citenamefont {Zhao}(2007)}]{zhao2007cultural}%
  \BibitemOpen
  \bibfield  {author} {\bibinfo {author} {\bibfnamefont {Y.}~\bibnamefont
  {Zhao}},\ }\href@noop {} {\bibfield  {journal} {\bibinfo  {journal}
  {Intercultural Communication Studies}\ }\textbf {\bibinfo {volume} {16}},\
  \bibinfo {pages} {129} (\bibinfo {year} {2007})}\BibitemShut {NoStop}%
\bibitem [{\citenamefont {Hu}(2002)}]{hu2002potential}%
  \BibitemOpen
  \bibfield  {author} {\bibinfo {author} {\bibfnamefont {G.}~\bibnamefont
  {Hu}},\ }\href@noop {} {\bibfield  {journal} {\bibinfo  {journal} {Language
  culture and curriculum}\ }\textbf {\bibinfo {volume} {15}},\ \bibinfo {pages}
  {93} (\bibinfo {year} {2002})}\BibitemShut {NoStop}%
\bibitem [{\citenamefont {Wilkinson}\ and\ \citenamefont
  {Olliver-Gray}(2006)}]{wilkinson2006significance}%
  \BibitemOpen
  \bibfield  {author} {\bibinfo {author} {\bibfnamefont {L.}~\bibnamefont
  {Wilkinson}}\ and\ \bibinfo {author} {\bibfnamefont {Y.}~\bibnamefont
  {Olliver-Gray}},\ }\href@noop {} {\bibfield  {journal} {\bibinfo  {journal}
  {Psychologia}\ }\textbf {\bibinfo {volume} {49}},\ \bibinfo {pages} {74}
  (\bibinfo {year} {2006})}\BibitemShut {NoStop}%
\bibitem [{\citenamefont {Rachel~Zhou}\ \emph {et~al.}(2005)\citenamefont
  {Rachel~Zhou}, \citenamefont {Knoke},\ and\ \citenamefont
  {Sakamoto}}]{rachel2005rethinking}%
  \BibitemOpen
  \bibfield  {author} {\bibinfo {author} {\bibfnamefont {Y.}~\bibnamefont
  {Rachel~Zhou}}, \bibinfo {author} {\bibfnamefont {D.}~\bibnamefont {Knoke}},
  \ and\ \bibinfo {author} {\bibfnamefont {I.}~\bibnamefont {Sakamoto}},\
  }\href@noop {} {\bibfield  {journal} {\bibinfo  {journal} {International
  Journal of Inclusive Education}\ }\textbf {\bibinfo {volume} {9}},\ \bibinfo
  {pages} {287} (\bibinfo {year} {2005})}\BibitemShut {NoStop}%
\bibitem [{\citenamefont {Wan}(1999)}]{wan1999learning}%
  \BibitemOpen
  \bibfield  {author} {\bibinfo {author} {\bibfnamefont {G.}~\bibnamefont
  {Wan}},\ }\href@noop {} {\  (\bibinfo {year} {1999})}\BibitemShut {NoStop}%
\bibitem [{\citenamefont {Beaver}\ and\ \citenamefont
  {Tuck}(1998)}]{beaver1998adjustment}%
  \BibitemOpen
  \bibfield  {author} {\bibinfo {author} {\bibfnamefont {B.}~\bibnamefont
  {Beaver}}\ and\ \bibinfo {author} {\bibfnamefont {B.}~\bibnamefont {Tuck}},\
  }\href@noop {} {\bibfield  {journal} {\bibinfo  {journal} {New Zealand
  Journal of Educational Studies}\ } (\bibinfo {year} {1998})}\BibitemShut
  {NoStop}%
\bibitem [{\citenamefont {Byars-Winston}\ \emph {et~al.}(2015)\citenamefont
  {Byars-Winston}, \citenamefont {Branchaw}, \citenamefont {Pfund},
  \citenamefont {Leverett},\ and\ \citenamefont
  {Newton}}]{byars2015culturally}%
  \BibitemOpen
  \bibfield  {author} {\bibinfo {author} {\bibfnamefont {A.~M.}\ \bibnamefont
  {Byars-Winston}}, \bibinfo {author} {\bibfnamefont {J.}~\bibnamefont
  {Branchaw}}, \bibinfo {author} {\bibfnamefont {C.}~\bibnamefont {Pfund}},
  \bibinfo {author} {\bibfnamefont {P.}~\bibnamefont {Leverett}}, \ and\
  \bibinfo {author} {\bibfnamefont {J.}~\bibnamefont {Newton}},\ }\href@noop {}
  {\bibfield  {journal} {\bibinfo  {journal} {International journal of science
  education}\ }\textbf {\bibinfo {volume} {37}},\ \bibinfo {pages} {2533}
  (\bibinfo {year} {2015})}\BibitemShut {NoStop}%
\bibitem [{\citenamefont {Chemers}\ \emph {et~al.}(2011)\citenamefont
  {Chemers}, \citenamefont {Zurbriggen}, \citenamefont {Syed}, \citenamefont
  {Goza},\ and\ \citenamefont {Bearman}}]{chemers2011role}%
  \BibitemOpen
  \bibfield  {author} {\bibinfo {author} {\bibfnamefont {M.~M.}\ \bibnamefont
  {Chemers}}, \bibinfo {author} {\bibfnamefont {E.~L.}\ \bibnamefont
  {Zurbriggen}}, \bibinfo {author} {\bibfnamefont {M.}~\bibnamefont {Syed}},
  \bibinfo {author} {\bibfnamefont {B.~K.}\ \bibnamefont {Goza}}, \ and\
  \bibinfo {author} {\bibfnamefont {S.}~\bibnamefont {Bearman}},\ }\href@noop
  {} {\bibfield  {journal} {\bibinfo  {journal} {Journal of Social Issues}\
  }\textbf {\bibinfo {volume} {67}},\ \bibinfo {pages} {469} (\bibinfo {year}
  {2011})}\BibitemShut {NoStop}%
\bibitem [{\citenamefont {Russell}\ \emph
  {et~al.}(2007{\natexlab{b}})\citenamefont {Russell}, \citenamefont
  {Hancock},\ and\ \citenamefont {McCullough}}]{russell_pipeline_2007}%
  \BibitemOpen
  \bibfield  {author} {\bibinfo {author} {\bibfnamefont {S.~H.}\ \bibnamefont
  {Russell}}, \bibinfo {author} {\bibfnamefont {M.~P.}\ \bibnamefont
  {Hancock}}, \ and\ \bibinfo {author} {\bibfnamefont {J.}~\bibnamefont
  {McCullough}},\ }\href@noop {} {\bibfield  {journal} {\bibinfo  {journal}
  {Science}\ }\textbf {\bibinfo {volume} {316}},\ \bibinfo {pages} {548}
  (\bibinfo {year} {2007}{\natexlab{b}})}\BibitemShut {NoStop}%
\bibitem [{\citenamefont {Engestr{\"o}m}(2014)}]{engestrom_learning_2014}%
  \BibitemOpen
  \bibfield  {author} {\bibinfo {author} {\bibfnamefont {Y.}~\bibnamefont
  {Engestr{\"o}m}},\ }\href {\doibase 10.1017/CBO9781139814744} {\emph
  {\bibinfo {title} {Learning by Expanding: An Activity-Theoretical Approach to
  Developmental Research}}}\ (\bibinfo  {publisher} {Cambridge University
  Press},\ \bibinfo {address} {Cambridge},\ \bibinfo {year} {2014})\BibitemShut
  {NoStop}%
\bibitem [{\citenamefont {Engestr{\"o}m}(1999)}]{engestrom_expansive_1999}%
  \BibitemOpen
  \bibfield  {author} {\bibinfo {author} {\bibfnamefont {Y.}~\bibnamefont
  {Engestr{\"o}m}},\ }\href@noop {} {\bibfield  {journal} {\bibinfo  {journal}
  {Comput. Support. Coop. Work.}\ }\textbf {\bibinfo {volume} {8}},\ \bibinfo
  {pages} {63} (\bibinfo {year} {1999})}\BibitemShut {NoStop}%
\bibitem [{\citenamefont {Gronn}(2002)}]{gronn_distributed_2002}%
  \BibitemOpen
  \bibfield  {author} {\bibinfo {author} {\bibfnamefont {P.}~\bibnamefont
  {Gronn}},\ }\href@noop {} {\bibfield  {journal} {\bibinfo  {journal}
  {Leadersh. Q.}\ }\textbf {\bibinfo {volume} {13}},\ \bibinfo {pages} {423}
  (\bibinfo {year} {2002})}\BibitemShut {NoStop}%
\bibitem [{\citenamefont {Marton}\ and\ \citenamefont
  {Booth}(1997{\natexlab{b}})}]{marton_learning_1997}%
  \BibitemOpen
  \bibfield  {author} {\bibinfo {author} {\bibfnamefont {F.}~\bibnamefont
  {Marton}}\ and\ \bibinfo {author} {\bibfnamefont {S.~A.}\ \bibnamefont
  {Booth}},\ }\href@noop {} {\emph {\bibinfo {title} {Learning and
  Awareness}}}\ (\bibinfo  {publisher} {Psychology Press},\ \bibinfo {year}
  {1997})\BibitemShut {NoStop}%
\bibitem [{\citenamefont {Sin}(2010)}]{sin_considerations_2010}%
  \BibitemOpen
  \bibfield  {author} {\bibinfo {author} {\bibfnamefont {S.}~\bibnamefont
  {Sin}},\ }\href@noop {} {\bibfield  {journal} {\bibinfo  {journal} {Int. J.
  Qual. Methods}\ } (\bibinfo {year} {2010})}\BibitemShut {NoStop}%
\bibitem [{\citenamefont {Booth}(1992)}]{booth_learning_1992}%
  \BibitemOpen
  \bibfield  {author} {\bibinfo {author} {\bibfnamefont {S.~A.}\ \bibnamefont
  {Booth}},\ }\href@noop {} {\emph {\bibinfo {title} {Learning to program: a
  phenomenographic perspective}}}\ (\bibinfo  {publisher} {G\"oteborg studies
  in educational sciences--Acta universitatis Gothoburgensis},\ \bibinfo {year}
  {1992})\BibitemShut {NoStop}%
\bibitem [{\citenamefont {Marton}(1981)}]{marton_phenomenography_1981}%
  \BibitemOpen
  \bibfield  {author} {\bibinfo {author} {\bibfnamefont {F.}~\bibnamefont
  {Marton}},\ }\href {\doibase https://doi.org/10.1007/BF00132516} {\bibfield
  {journal} {\bibinfo  {journal} {Instr. Sci.}\ }\textbf {\bibinfo {volume}
  {10}},\ \bibinfo {pages} {177} (\bibinfo {year} {1981})}\BibitemShut
  {NoStop}%
\bibitem [{\citenamefont {Marton}(1986)}]{marton_phenomenographyresearch_1986}%
  \BibitemOpen
  \bibfield  {author} {\bibinfo {author} {\bibfnamefont {F.}~\bibnamefont
  {Marton}},\ }\href {https://www.jstor.org/stable/42589189} {\bibfield
  {journal} {\bibinfo  {journal} {Journal of Thought}\ }\textbf {\bibinfo
  {volume} {21}},\ \bibinfo {pages} {28} (\bibinfo {year} {1986})},\ \bibinfo
  {note} {publisher: Caddo Gap Press}\BibitemShut {NoStop}%
\bibitem [{\citenamefont {Mendoza}\ \emph {et~al.}(2015)\citenamefont
  {Mendoza}, \citenamefont {Richard},\ and\ \citenamefont
  {Wickliff}}]{mendoza_enculturation_2015}%
  \BibitemOpen
  \bibfield  {author} {\bibinfo {author} {\bibfnamefont {N.}~\bibnamefont
  {Mendoza}}, \bibinfo {author} {\bibfnamefont {J.}~\bibnamefont {Richard}}, \
  and\ \bibinfo {author} {\bibfnamefont {T.~D.}\ \bibnamefont {Wickliff}},\
  }in\ \href@noop {} {\emph {\bibinfo {booktitle} {Paper Presented at the 7th
  First Year Engineering Experiences Conference}}}\ (\bibinfo {address}
  {Roanoke, VA.},\ \bibinfo {year} {2015})\BibitemShut {NoStop}%
\bibitem [{\citenamefont {Prior}\ and\ \citenamefont
  {Bilbro}(2012)}]{prior_academic_2012}%
  \BibitemOpen
  \bibfield  {author} {\bibinfo {author} {\bibfnamefont {P.}~\bibnamefont
  {Prior}}\ and\ \bibinfo {author} {\bibfnamefont {R.}~\bibnamefont {Bilbro}},\
  }in\ \href@noop {} {\emph {\bibinfo {booktitle} {University Writing: Selves
  and Texts in Academic Societies}}},\ \bibinfo {series and number} {Studies in
  Writing},\ \bibinfo {editor} {edited by\ \bibinfo {editor} {\bibfnamefont
  {M.}~\bibnamefont {Castello}}\ and\ \bibinfo {editor} {\bibfnamefont
  {C.}~\bibnamefont {Donahue}}}\ (\bibinfo  {publisher} {Emerald},\ \bibinfo
  {year} {2012})\ pp.\ \bibinfo {pages} {19--31}\BibitemShut {NoStop}%
\bibitem [{\citenamefont {Zydney}\ \emph {et~al.}(2002)\citenamefont {Zydney},
  \citenamefont {Bennett}, \citenamefont {Shahid},\ and\ \citenamefont
  {Bauer}}]{zydney_impact_2002}%
  \BibitemOpen
  \bibfield  {author} {\bibinfo {author} {\bibfnamefont {A.~L.}\ \bibnamefont
  {Zydney}}, \bibinfo {author} {\bibfnamefont {J.~S.}\ \bibnamefont {Bennett}},
  \bibinfo {author} {\bibfnamefont {A.}~\bibnamefont {Shahid}}, \ and\ \bibinfo
  {author} {\bibfnamefont {K.~W.}\ \bibnamefont {Bauer}},\ }\href@noop {}
  {\bibfield  {journal} {\bibinfo  {journal} {J. Eng. Educ.}\ }\textbf
  {\bibinfo {volume} {91}},\ \bibinfo {pages} {151} (\bibinfo {year}
  {2002})}\BibitemShut {NoStop}%
\bibitem [{\citenamefont {Gonzalez-Ramirez}\ \emph {et~al.}(2021)\citenamefont
  {Gonzalez-Ramirez}, \citenamefont {Mulqueen}, \citenamefont {Zealand},
  \citenamefont {Silverstein}, \citenamefont {Mulqueen},\ and\ \citenamefont
  {BuShell}}]{gonzalez2021emergency}%
  \BibitemOpen
  \bibfield  {author} {\bibinfo {author} {\bibfnamefont {J.}~\bibnamefont
  {Gonzalez-Ramirez}}, \bibinfo {author} {\bibfnamefont {K.}~\bibnamefont
  {Mulqueen}}, \bibinfo {author} {\bibfnamefont {R.}~\bibnamefont {Zealand}},
  \bibinfo {author} {\bibfnamefont {S.}~\bibnamefont {Silverstein}}, \bibinfo
  {author} {\bibfnamefont {C.}~\bibnamefont {Mulqueen}}, \ and\ \bibinfo
  {author} {\bibfnamefont {S.}~\bibnamefont {BuShell}},\ }\href@noop {}
  {\bibfield  {journal} {\bibinfo  {journal} {College Student Journal}\
  }\textbf {\bibinfo {volume} {55}},\ \bibinfo {pages} {29} (\bibinfo {year}
  {2021})}\BibitemShut {NoStop}%
\bibitem [{\citenamefont {Yamagata-Lynch}(2010)}]{yamagata2010activity}%
  \BibitemOpen
  \bibfield  {author} {\bibinfo {author} {\bibfnamefont {L.~C.}\ \bibnamefont
  {Yamagata-Lynch}},\ }\href@noop {} {\emph {\bibinfo {title} {Activity systems
  analysis methods: Understanding complex learning environments}}}\ (\bibinfo
  {publisher} {Springer Science \& Business Media},\ \bibinfo {year}
  {2010})\BibitemShut {NoStop}%
\bibitem [{\citenamefont {Zhang}\ \emph {et~al.}(2020)\citenamefont {Zhang},
  \citenamefont {Wang}, \citenamefont {Yang},\ and\ \citenamefont
  {Wang}}]{zhang2020suspending}%
  \BibitemOpen
  \bibfield  {author} {\bibinfo {author} {\bibfnamefont {W.}~\bibnamefont
  {Zhang}}, \bibinfo {author} {\bibfnamefont {Y.}~\bibnamefont {Wang}},
  \bibinfo {author} {\bibfnamefont {L.}~\bibnamefont {Yang}}, \ and\ \bibinfo
  {author} {\bibfnamefont {C.}~\bibnamefont {Wang}},\ }\href@noop {} {\enquote
  {\bibinfo {title} {Suspending classes without stopping learning: China’s
  education emergency management policy in the covid-19 outbreak},}\ }
  (\bibinfo {year} {2020})\BibitemShut {NoStop}%
\bibitem [{\citenamefont {Nambiar}(2020)}]{nambiar2020impact}%
  \BibitemOpen
  \bibfield  {author} {\bibinfo {author} {\bibfnamefont {D.}~\bibnamefont
  {Nambiar}},\ }\href@noop {} {\bibfield  {journal} {\bibinfo  {journal} {The
  International Journal of Indian Psychology}\ }\textbf {\bibinfo {volume}
  {8}},\ \bibinfo {pages} {783} (\bibinfo {year} {2020})}\BibitemShut {NoStop}%
\bibitem [{\citenamefont {Yaffe}\ \emph {et~al.}(2014)\citenamefont {Yaffe},
  \citenamefont {Bender},\ and\ \citenamefont {Sechrest}}]{yaffe_how_2014}%
  \BibitemOpen
  \bibfield  {author} {\bibinfo {author} {\bibfnamefont {K.}~\bibnamefont
  {Yaffe}}, \bibinfo {author} {\bibfnamefont {C.}~\bibnamefont {Bender}}, \
  and\ \bibinfo {author} {\bibfnamefont {L.}~\bibnamefont {Sechrest}},\
  }\href@noop {} {\bibfield  {journal} {\bibinfo  {journal} {J. Coll. Sci.
  Teach.}\ }\textbf {\bibinfo {volume} {44}},\ \bibinfo {pages} {25} (\bibinfo
  {year} {2014})},\ \bibinfo {note} {publisher: National Science Teachers
  Association}\BibitemShut {NoStop}%
\bibitem [{\citenamefont {Carpi}\ \emph {et~al.}(2017)\citenamefont {Carpi},
  \citenamefont {Ronan}, \citenamefont {Falconer},\ and\ \citenamefont
  {Lents}}]{carpi_cultivating_2017}%
  \BibitemOpen
  \bibfield  {author} {\bibinfo {author} {\bibfnamefont {A.}~\bibnamefont
  {Carpi}}, \bibinfo {author} {\bibfnamefont {D.~M.}\ \bibnamefont {Ronan}},
  \bibinfo {author} {\bibfnamefont {H.~M.}\ \bibnamefont {Falconer}}, \ and\
  \bibinfo {author} {\bibfnamefont {N.~H.}\ \bibnamefont {Lents}},\ }\href
  {\doibase https://doi.org/10.1002/tea.21341} {\bibfield  {journal} {\bibinfo
  {journal} {J. Res. Sci. Teach.}\ }\textbf {\bibinfo {volume} {54}},\ \bibinfo
  {pages} {169} (\bibinfo {year} {2017})}\BibitemShut {NoStop}%
\bibitem [{\citenamefont {Mabrouk}\ and\ \citenamefont
  {Peters}(2000)}]{mabrouk_student_2000}%
  \BibitemOpen
  \bibfield  {author} {\bibinfo {author} {\bibfnamefont {P.~A.}\ \bibnamefont
  {Mabrouk}}\ and\ \bibinfo {author} {\bibfnamefont {K.}~\bibnamefont
  {Peters}},\ }\href@noop {} {\bibfield  {journal} {\bibinfo  {journal}
  {Council on Undergraduate Research (CUR) Quarterly}\ ,\ \bibinfo {pages}
  {20}} (\bibinfo {year} {2000})}\BibitemShut {NoStop}%
\bibitem [{\citenamefont {l.~Gafney}(2001)}]{gafney_impact_2001}%
  \BibitemOpen
  \bibfield  {author} {\bibinfo {author} {\bibnamefont {l.~Gafney}},\
  }\href@noop {} {\bibfield  {journal} {\bibinfo  {journal} {Council on
  Undergraduate Research (CUR) Quarterly}\ }\textbf {\bibinfo {volume} {21}},\
  \bibinfo {pages} {172} (\bibinfo {year} {2001})}\BibitemShut {NoStop}%
\bibitem [{\citenamefont {Kardash}\ \emph {et~al.}(2008)\citenamefont
  {Kardash}, \citenamefont {Wallace},\ and\ \citenamefont
  {Blockus}}]{kardash_undergraduate_2008}%
  \BibitemOpen
  \bibfield  {author} {\bibinfo {author} {\bibfnamefont {C.~M.}\ \bibnamefont
  {Kardash}}, \bibinfo {author} {\bibfnamefont {M.}~\bibnamefont {Wallace}}, \
  and\ \bibinfo {author} {\bibfnamefont {L.}~\bibnamefont {Blockus}},\ }in\
  \href@noop {} {\emph {\bibinfo {booktitle} {Creating Effective Undergraduate
  Research Programs in Science}}}\ (\bibinfo  {publisher} {Teachers College
  Press},\ \bibinfo {address} {New York, NY, US},\ \bibinfo {year} {2008})\
  pp.\ \bibinfo {pages} {191--205}\BibitemShut {NoStop}%
\end{thebibliography}%

\end{document}